\documentclass[epj,nopacs]{svjour}

\usepackage{amssymb}
\usepackage{feynarts}
\usepackage{epsfig}
\usepackage{cancel}

\def\URLtilde{\lower0.2em\hbox{$\tilde{\phantom{a}}$}}

\def\beq{\begin{equation}}
\def\eeq{\end{equation}}

\def\beqn{\begin{eqnarray}}
\def\eeqn{\end{eqnarray}}

\def\mhat{\widehat{m}}
\def\slashchar#1{\setbox0=\hbox{$#1$}           
   \dimen0=\wd0                                 
   \setbox1=\hbox{/} \dimen1=\wd1               
   \ifdim\dimen0>\dimen1                        
      \rlap{\hbox to \dimen0{\hfil/\hfil}}#1
   \else                                        
      \rlap{\hbox to \dimen1{\hfil$#1$\hfil}}/
       \fi}

\newcommand{\ud}{\mathrm{d}}

\begin{document}

\title{Spin Correlations in Decay Chains Involving $W$ Bosons\thanks{Work supported by the
  UK Particle Physics and Astronomy Research Council.}}

\author{Jennifer M.\ Smillie\inst{1}} 

\institute{Cavendish Laboratory, University of Cambridge, 
  JJ Thomson Avenue, Cambridge CB3 0HE, U.K. \\
  \email{smillie@hep.phy.cam.ac.uk}}

\date{Received: date/ Revised version: date}

\abstract{We study the extent to which spin assignments of new particles produced at the
  LHC can be deduced in the decay of a scalar or fermion $C$ into a new stable (or
  quasi-stable) particle $A$ through the chain $C\to B^{\pm} q, B^{\pm}\to A W^{\pm},
  W^{\pm}\to \ell^{\pm} \nu_{\ell}$ where $\ell=e,\mu$.  All possible spin assignments of
  the particles $A$ and $B$ are considered.  Explicit invariant mass distributions of the
  quark and lepton are given for each set of spins, valid for all masses.  We also
  construct the asymmetry between the chains with a $W^-$ and those with a $W^+$.  The
  Kullback-Leibler distance between the distributions is then calculated to give a
  quantitative measure of our ability to distinguish the different spin assignments.}

\maketitle

\section{Introduction}
\label{sec:introduction}

While the Standard Model (SM) has been remarkably successful to date, new physics is
expected around the TeV scale, for example to cancel the large contributions to the Higgs
mass thereby solving the Hierarchy problem.  Whatever form this new physics takes, we
expect to find new particles.  The issue of deducing the spin of these new particles from
experimental data has become increasingly important with the rise in popularity of
supersymmetric (SUSY) extensions to the SM.  These models assign to SM partners different
spin to that of the corresponding SM particle.  

This is in contrast to another possible SM extension, Universal Extra Dimensions (UED)
\cite{Appelquist:2000nn} where each SM partner has the same spin as its SM counterpart.
In these models all fields propagate into at least one extra dimension forming
Kaluza-Klein towers of new particles with increasing mass but otherwise identical quantum
numbers.  From this construction, a typical UED mass spectrum is very degenerate.  There
are also other possible extensions to the Standard Model such as Little Higgs Models
\cite{Arkani-Hamed:2001nc} where the Higgs field is a pseudo Nambu Goldstone boson from a
broken symmetry group.  These models often feature different spin assignments to new
particles, such as new scalars without a direct SM counterpart.

Often, studies of spin are considered in the context of a linear electron collider.
However, Barr \cite{Barr:2004ze} (see also \cite{Goto:2004cp}) showed that it was possible
to deduce such information at the Large Hadron Collider (LHC).  He demonstrated that one
could distinguish between the case where particles had SUSY spin allocations and where the
particles were all effectively spinless.  This work was extended in
\cite{Smillie:2005ar,Battaglia:2005zf,Datta:2005zs,Datta:2005vx,Barr:2005dz,Alves:2006df}
to demonstrate that spin studies were a useful tool to distinguish between SUSY and UED.
(It was first pointed out in \cite{Cheng:2002ab} that these two models could mimic each
other.)  Recently \cite{Athanasiou:2006ef}, the technique was extended to cover all
possible spin assignments in the cascade decay of a quark partner via
opposite-sign-same-flavour (OSSF) leptons.  This had much wider applications as it was no
longer constrained to spin effects only in the Minimal Supersymmetric Standard Model
(MSSM) and UED.  A similar study had previously been applied in \cite{Meade:2006dw} to the
pair production of top quark partners each decaying straight to a top and a stable particle.

These studies have concentrated on the quark partner cascade decay (or gluino decay
leading to this), top partner production and Drell-Yan production of lepton pairs and
their subsequent decay.  Here we study the electroweak decay of a quark partner via a $W$
boson decaying leptonically.  In the MSSM, this decay chain often has a higher branching
ratio than the cascade decay via a $\widetilde{\chi}^0_2$ which is more frequently
studied. In \cite{Wang:2006hk}, it is suggested that this could be the most promising
channel for spin discrimination.  Here, we consider all possible spin assignments so as
not to constrain ourselves to a particular model.  We assume that these chains have been
identified, and that the masses of the particles involved are known.  The spin
correlations in the chain depend on the charge of the $W$ boson, so we consider the two
charge assignments separately.

In section \ref{sec:spin-assignments}, we discuss all possible spin assignments in the
decay chain and the resulting matrix elements.  In section \ref{sec:spin-correlations},
spin correlations are discussed in terms of the invariant mass distributions of the quark
and lepton.  The full analytical formulae valid for any mass spectrum are calculated.  We
then form an asymmetry between the chains with a $W^-$ and the chains with a $W^+$.  These
distributions are plotted and discussed.  In section \ref{sec:model-discrimination}, we
quantify the results of the previous section using the Kullback-Leibler distance method
introduced in \cite{Athanasiou:2006ef}.  This gives a lower limit on the number of events
required to discriminate between any two of the spin allocations at a given level of
confidence.  These lower limits do not include background or detector effects as these
will vary between experiments and we wish to remain general.  They do however provide a
handle on the feasibility of distinguishing two particular curves.  This analysis is
applied to the observable processes individually, and then combined.  The conclusions are
in section \ref{sec:conclusions} before the more lengthy formulae and a discussion of
higher derivative vertices which are in the appendix.

\section{Spin assignments}
\label{sec:spin-assignments}

\begin{figure}[!h]
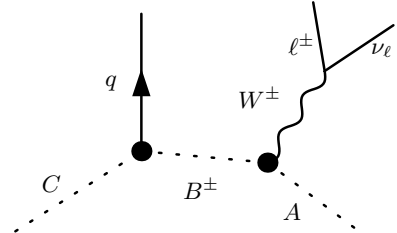

  \unitlength=0.5bp%
  \centering
  \vspace{-2cm}
  \hspace{1.5cm}
  \begin{feynartspicture}(432,380)(1,1)    
    \FADiagram{}
    \FAProp(0.,5.)(5.5,8.5)(0.,){/GhostDash}{0}
    \FALabel(2.,6.7)[br]{$C$}
    \FAProp(5.5,14.5)(5.5,8.5)(0.,){/Straight}{-1}
    \FALabel(4.43,11.5)[r]{$q$}
    \FAProp(13.5,12.)(11.,8.)(0.,){/Sine}{0}
    \FALabel(11.6,10.3995)[br]{$W^{\pm}$}
    \FAProp(5.5,8.5)(11.,8.)(0.,){/GhostDash}{0}
    \FALabel(8.10967,7.3)[t]{$B^{\pm}$}
    \FAProp(11.,8.)(15.,5.)(0.,){/GhostDash}{0}
    \FALabel(12.461,6.2)[tr]{$A$}
    \FAProp(13.5,12.)(13.,15.)(0.,){/Straight}{0}
    \FALabel(12.5,12.7)[b]{$\ell^{\pm}$}
    \FAProp(13.5,12.)(16.5,14.)(0.,){/Straight}{0}
    \FALabel(16.,13.2)[t]{$\nu_{\ell}$}
    \FAVert(5.5,8.5){0}
    \FAVert(11.,8.){0}
  \end{feynartspicture}
  \vspace{-2cm}
  \caption{The decay chain under consideration.}
  \label{fig:decaychain}
\end{figure}

We will consider the decay of a heavy colour-triplet scalar or fermion $C$ of the form $C
\to B^{\pm} q$, $B^{\pm}\to A W^{\pm}$, $W^{\pm}\to \ell^{\pm} \nu_{\ell}$ (figure
\ref{fig:decaychain}), where $\ell=e,\mu$.  Chains like this can occur in SUSY, UED or
the Littlest Higgs model with T-parity.

We will assume that particle $A$ is a stable or long-lived heavy massive new particle and
that the masses of the new heavy particles $A,B$ and $C$ have all been measured.  All
possible spin configurations are listed in Table \ref{tab:spin-configs}, together with the
labels which will be used in the rest of the paper.

\begin{table}[!htbp]
  \centering
  \begin{tabular}{c|c|c|c}
    Label & C & B & A \\ \hline 
    SFF & Scalar & Fermion & Fermion \\
    FSS & Fermion & Scalar & Scalar \\
    FSV & Fermion & Scalar & Vector \\
    FVS & Fermion & Vector & Scalar \\
    FVV & Fermion & Vector & Vector \\
  \end{tabular}
  \caption{Possible spin configurations in the decay chain (figure \ref{fig:decaychain}).}
  \label{tab:spin-configs}
\end{table}

The SFF chain corresponds to SUSY spin assignments while FVV corresponds to the spin
assignments in a UED model, or a Littlest Higgs model with T-parity \cite{Cheng:2004yc}.
The other spin assignments correspond to non-minimal versions of these or other
models.  The UED masses derived from \cite{Cheng:2002iz} do not allow a decay chain of
this form to proceed for values of the compactification radius accessible at the LHC,
however, these were calculated under the assumption that the orbifold boundary kinetic
terms vanish at the cut-off scale.  This is not necessarily the case and for different
values of these parameters it is possible that the decay chain would still proceed.
Indeed, it is the freedom in choice of parameters which makes it unlikely that these
models could be distinguished by studying mass spectra alone.

It is necessary to make some assumptions about the structure of the vertices in the
chains, except for the SFF chain where these are well-determined in the MSSM.  We are not
concerned with overall numerical factors as the distributions are normalised to integrate
to 1.  When we consider the FSS, FSV, FVS and FVV chains, the $B$-$W$-$A$ vertex structure
is uniquely determined if we do not consider higher dimensional couplings like those
induced from loops.  In the FSS chain, it is of the form $(p-q)^{\mu}$ where $p$ and $q$
are the incoming momenta of the scalars; in the FSV and FVS chains it is of the form
$g^{\mu \nu}$, where $\mu$ is the index corresponding to the W and $\nu$ is the index
corresponding to the other vector particle ($A$ in FSV or $B$ in FVS), while in the FVV
chain the triple vector vertex takes the form
$g^{\mu\nu}(p_1-p_2)^{\rho}+g^{\nu\rho}(p_2-p_3)^{\mu}+g^{\rho\mu}(p_3-p_1)^{\nu}$.  These
are the structures considered here --- a discussion of possible alternatives is in appendix
\ref{sec:loop-induc-coupl}.

The structure of the $C$-$q$-$B$ vertex is not so well determined in these chains and in
principle contains a factor of $(1+a \gamma_5)$ where $a$ is an arbitrary constant.  In
the massless $q$ limit, the final distributions are in fact independent of $a$ for the FSS
and FSV chains, but this is not the case for FVS and FVV.  For these chains, where
necessary, the constant $a$ value has been taken to be $-1$, thereby forcing the $q$ to be
left-handed.  This value is justified because most models beyond the standard model have
two excitations for each fermion --- one coupling to the left-handed fermion and one
coupling to the right.  As they have the SM as a low energy limit, it is usually the one
associated with the left-handed fermion which undergoes decays of the type in figure
\ref{fig:decaychain} (especially for the light quarks we consider where left-right mixing
is expected to be small).  In particular, the FVV chain has the spin structure found in
UED where this is the case.

\section{Spin correlations}
\label{sec:spin-correlations}

In the chain, there are only two observable emitted particles, the quark (jet) and the
charged lepton.  This gives one observable invariant mass-squared:
$m_{q\ell}^2=(p_q+p_{\ell})^2$.  We define the angle $\theta$ to be the angle between the
quark and $A$ in the rest frame of $B$, and $\psi$ to be the angle between the lepton and
$A$ in the rest frame of the $W^{\pm}$.  We then define $\phi$ to be the angle between
these two planes.  Then,
\begin{eqnarray}
  \label{eq:mqmu}
 m_{q\ell}^2&=&\frac{1}{4X}m_B^2(1-X)\left(
  (1+Y-Z)(1-\cos\theta\cos\psi) \phantom{\sqrt{Y}} \right. \nonumber \\ && \;  +
  \sqrt{(1+Y-Z)^2-4Y}(\cos\theta-\cos\psi) \nonumber \\ && \; \left. -
  2\sqrt{Y}\sin\theta\sin\psi\cos\phi\right),
\end{eqnarray}
where the mass-squared ratios $X,Y,Z$ are $X=m_B^2/m_C^2$, $Y=m_W^2/m_B^2$ and
$Z=m_A^2/m_B^2$.  These must satisfy $\sqrt{Y}+\sqrt{Z}\le 1$ by energy conservation and
so the quantity in the square root is always non-negative.  The maximum value of
$m_{q\ell}^2$ is $\frac{1}{2X}m_B^2(1-X)((1+Y-Z)+\sqrt{(1+Y-Z)^2-4Y})$ which occurs when
$(\theta,\psi)=(0,\pi)$.

In order to keep a manageable expression, we define the scaled invariant mass as 
\begin{equation}
  \label{eq:mhat}
  \widehat{m}_{q\ell}^2=\frac{4X}{m_B^2(1-X)} m_{q\ell}^2, 
\end{equation}
which lies in the interval 
\begin{center}
$\left[0,2((1+Y-Z)+\sqrt{(1+Y-Z)^2-4Y})\right]$.
\end{center}

The analytical expressions valid for any particle masses are discussed in appendix
\ref{sec:analytical-formulae}, however, in order to plot the functions we must choose
values for the masses of $A,B$ and $C$.  If we consider this chain in a SUSY scenario, we
have the particle assignments given in table \ref{tab:sps1a}.  The masses given are those
at the Snowmass Benchmark points, SPS 1a, SPS 2 and SPS 9 \cite{Allanach:2002nj}.  SPS 1a
and SPS 2 were chosen as the points with the biggest difference in their spectrum, while the
AMSB point, SPS 9, was chosen as an example of a heavier chargino which allows for a
greater difference between the mass ratios $Y$ and $Z$.

\begin{table}[!htbp]
  \centering
  \begin{tabular}{|c|c|c|c|}
    \hline
    & $C$ & $B$ & $A$ \\ 
    & $\widetilde{u}_L$ & $\widetilde{\chi}_2^{\pm}$ & $\widetilde{\chi}_1^0$ \\ \hline
    SPS 1a &  537 & 378 & 96 \\ \hline
    SPS 2 & 1533 & 269 & 79 \\ \hline
    SPS 9 & 1237 & 876 & 175 \\ \hline
  \end{tabular}
  \caption{The mass spectra (in GeV) considered in this paper.}
  \label{tab:sps1a}
\end{table}

The spin correlations in the chain where the quark partner decays through a $W^-$ has
different spin correlations to that in which the quark partner decays through a $W^+$ as
one has a charged lepton, the other a charged anti-lepton (this sends $a\to -a$).  This
means we have two processes to consider:
\begin{itemize}
\item[] Process 1: \{$q$,$W$\} = \{$u$,$W^-$\} or \{$\bar{u}$,$W^+$\}
\item[] Process 2: \{$q$,$W$\} = \{$d$,$W^+$\} or \{$\bar{d}$,$W^-$\}.
\end{itemize}
Here $u$ stands for either an up or a charm quark and $d$ stands for a down or a strange
quark.  We do not include bottom and top quarks since $b$ and $t$ final states should be
distinguishable from those due to the lighter quarks.  We may then work in the massless
approximation.  For the FSS and FSV chains, processes 1 and 2 give the same distribution
as the scalar does not carry spin information down the chain.

\begin{figure}[!htbp]
  \centering
  \includegraphics[width=0.47\textwidth]{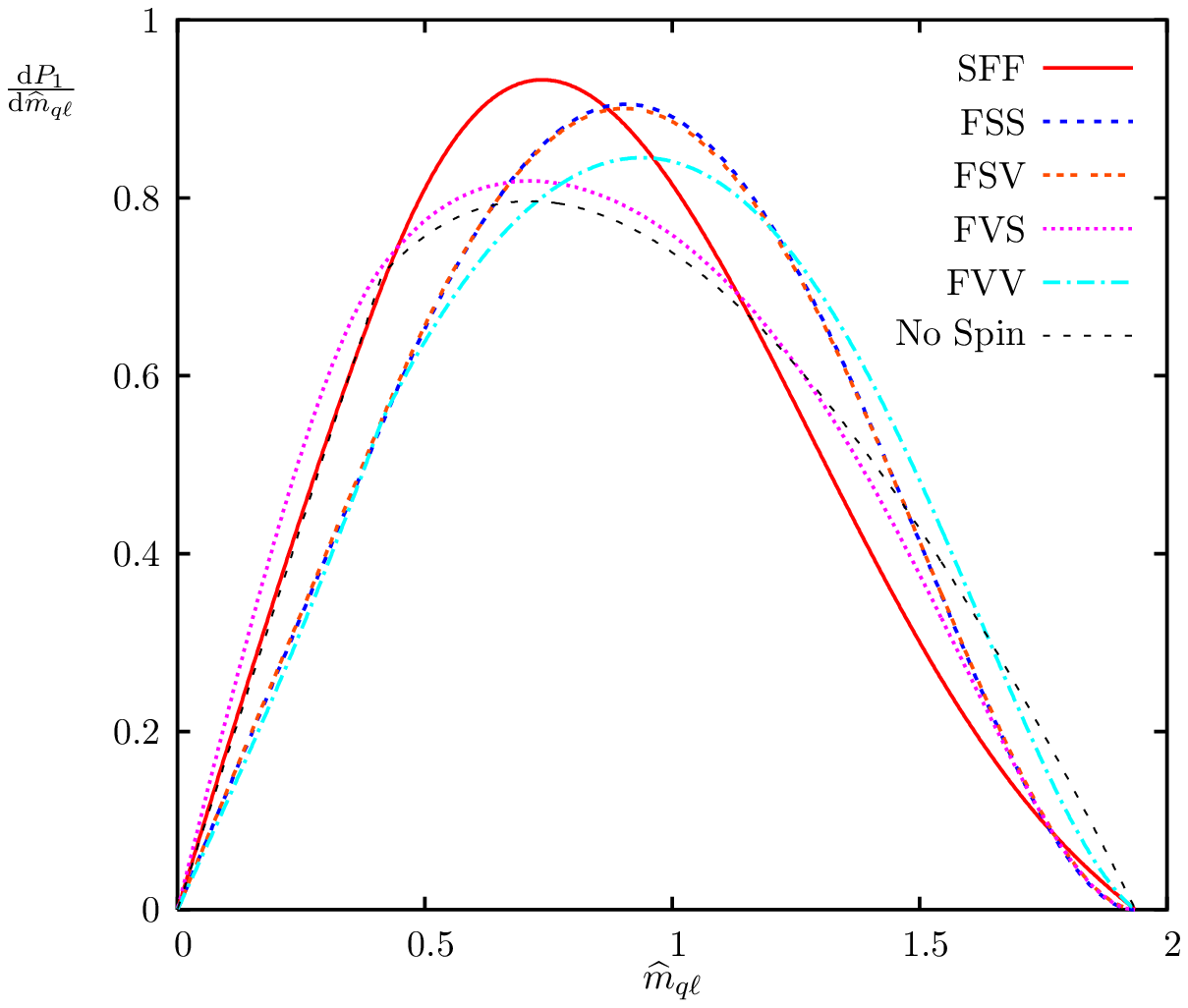}
  \includegraphics[width=0.47\textwidth]{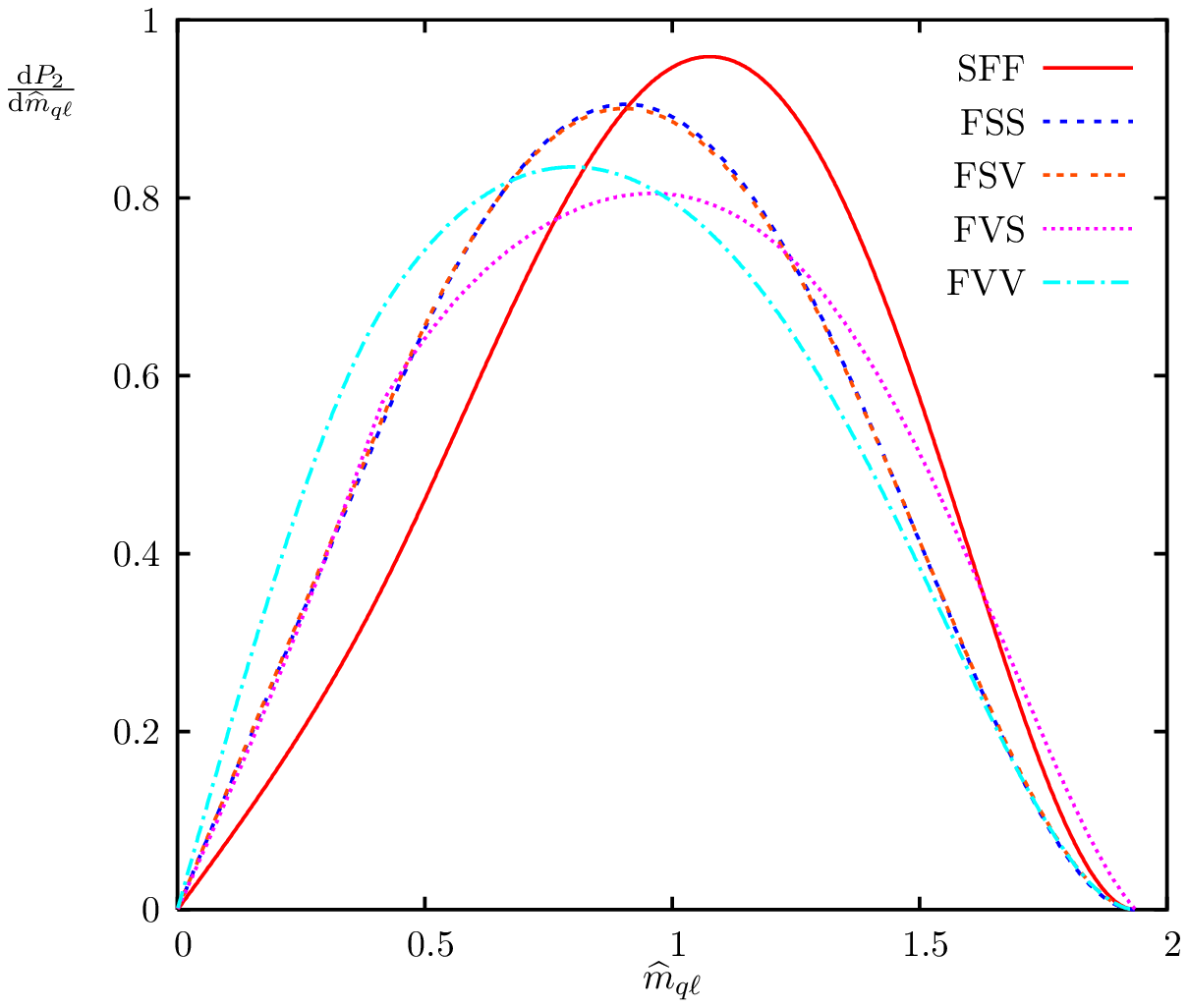}
  \caption{Invariant mass distributions for SPS 1a: Process 1 (top) and Process 2
    (bottom).} 
  \label{fig:P1and2}
\end{figure}

\begin{figure}[!htbp]
  \centering
  \includegraphics[width=0.47\textwidth]{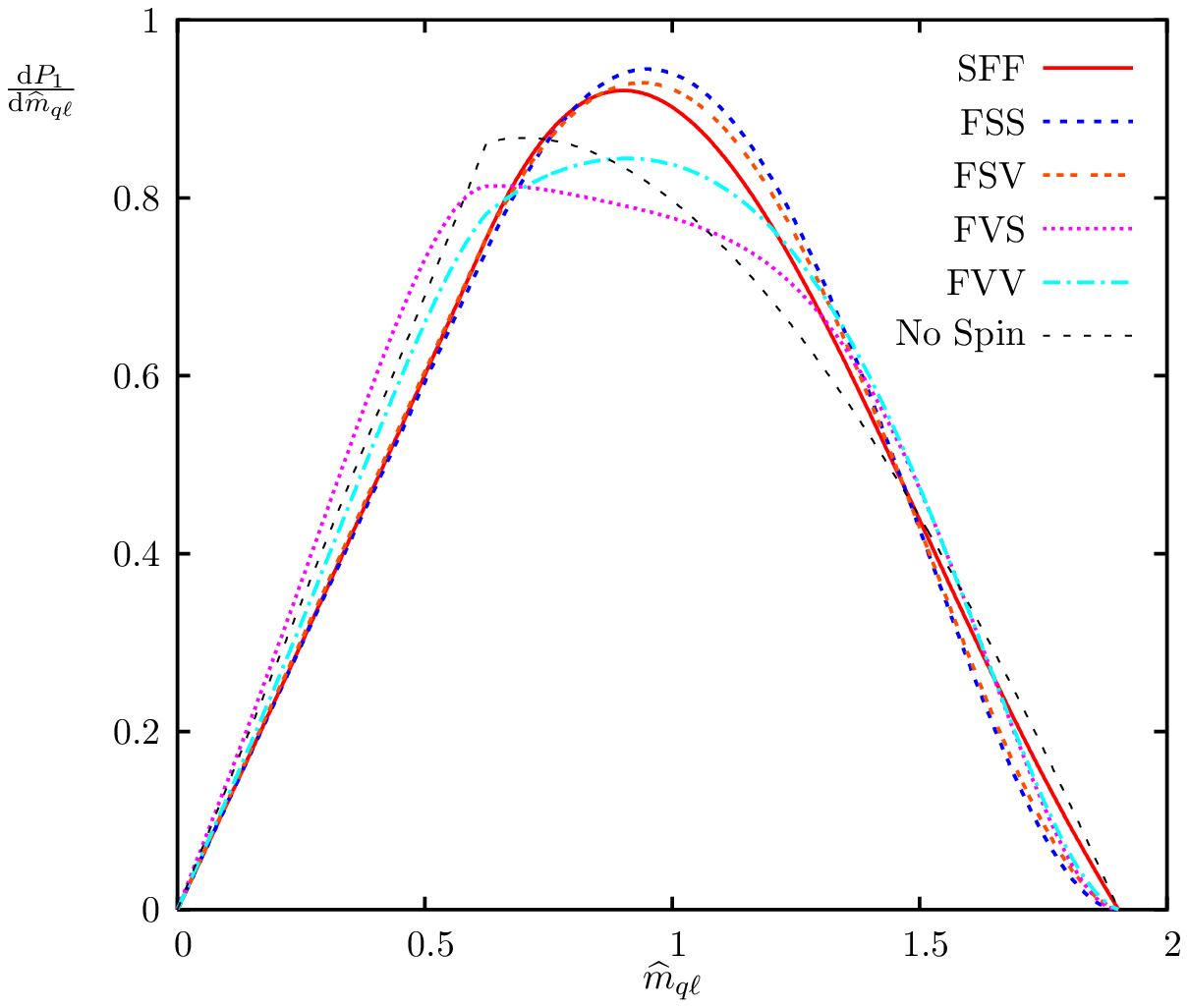}
  \includegraphics[width=0.47\textwidth]{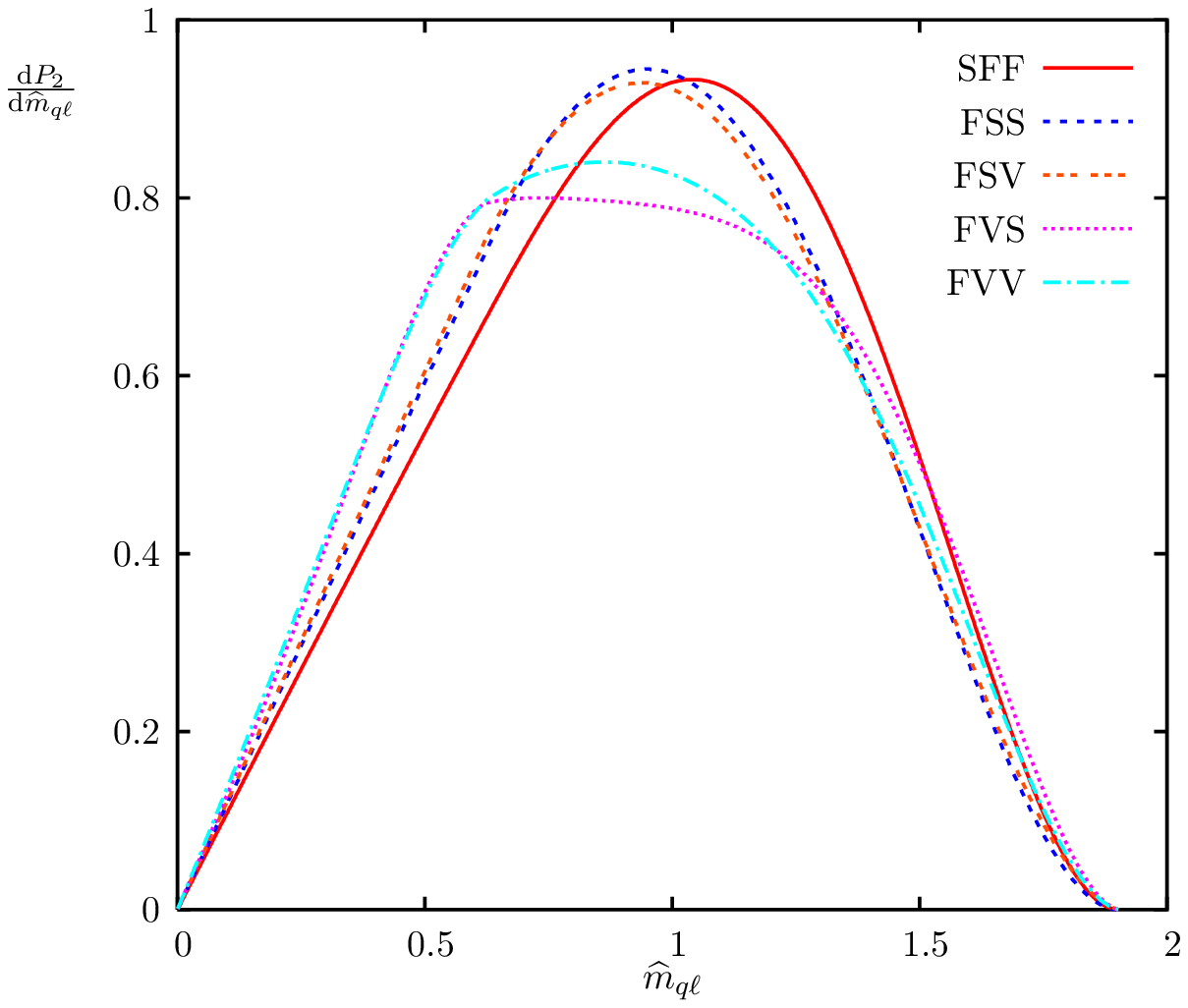}
  \caption{Invariant mass distributions for SPS 2: Process 1 (top) and Process 2 (bottom).} 
  \label{fig:SPS2}
\end{figure}

\begin{figure}[!htbp]
  \centering
  \includegraphics[width=0.47\textwidth]{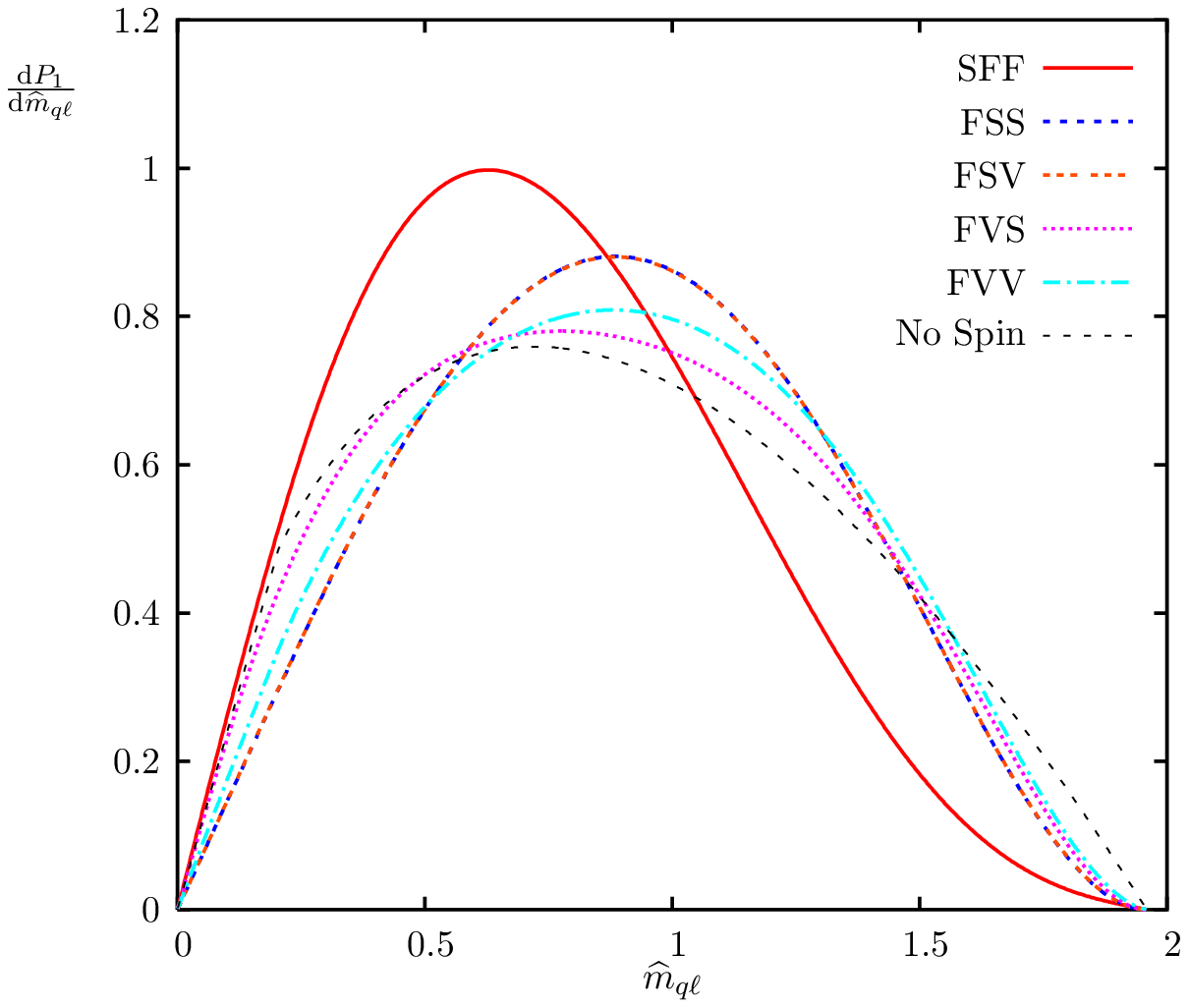}
  \includegraphics[width=0.47\textwidth]{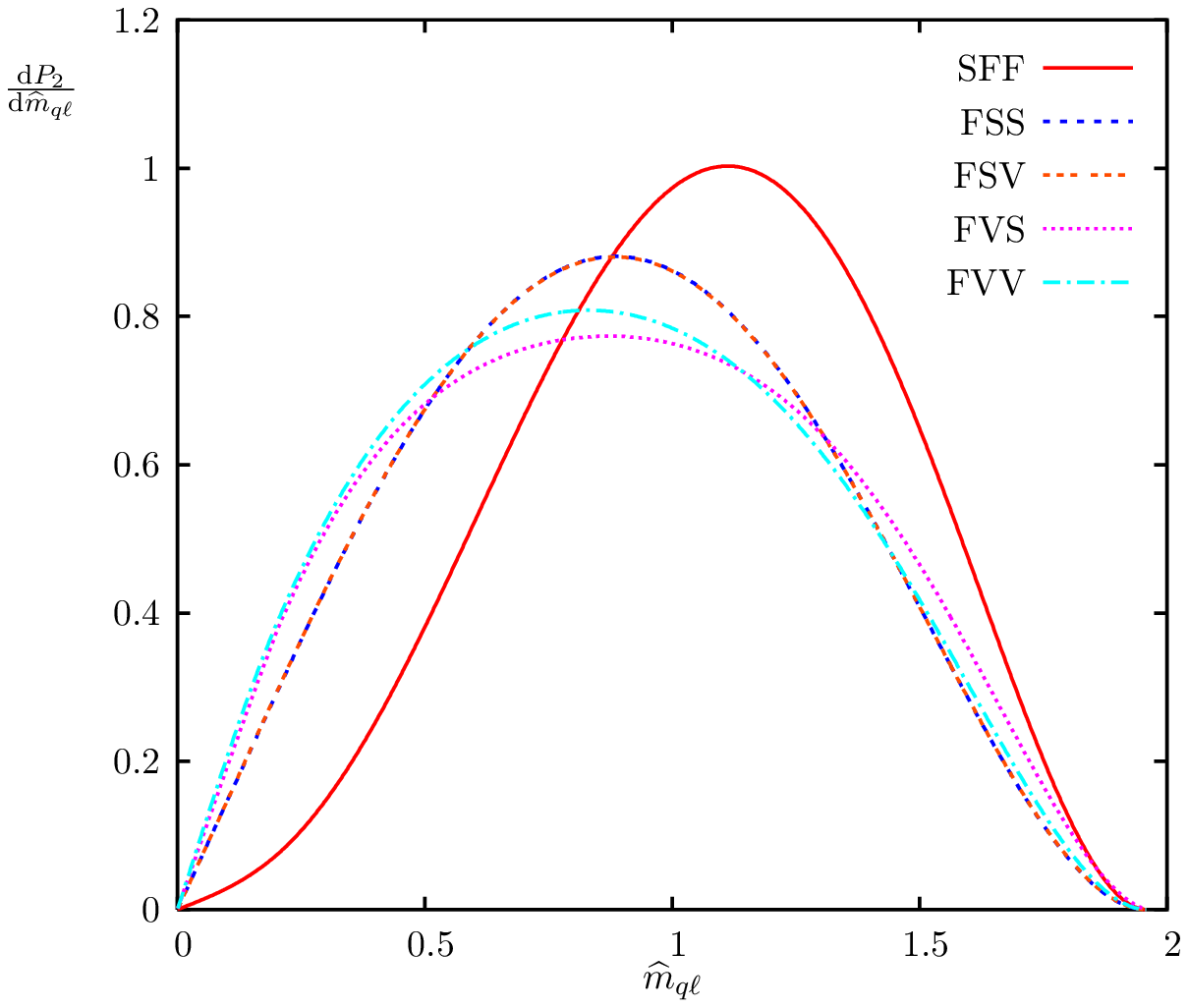}
  \caption{Invariant mass distributions for SPS 9: Process 1 (top) and Process 2 (bottom).} 
  \label{fig:SPS9}
\end{figure}

Figure \ref{fig:P1and2} shows the invariant mass-squared distributions for both processes
for the SPS 1a mass spectrum, for the five spin assignments given in Table
\ref{tab:spin-configs}.  Here, the distributions are plotted as $\ud P/\ud \mhat$
throughout, as opposed to $\ud P/\ud \mhat^2$ as was done in \cite{Athanasiou:2006ef}, as
the phase space is not flat in any such simple mapping of the invariant mass.  The phase
space curve (the case where all particles are treated as spinless) also depends on the
masses in the chain, and so is indicated on the $\ud P_1/\ud \mhat$ plot for each mass
spectrum (marked No Spin).
Figures \ref{fig:SPS2} and \ref{fig:SPS9} show the same distributions for the five spin
assignments, for the SPS 2 and SPS 9 mass spectra.

These plots show that all the curves have a similar overall shape, but with some
differences due to the different Lorentz structure.  The exact effect can be seen in the
equations in appendix \ref{sec:analytical-formulae}.  However, quantitative statements can
be made.  For example, the SFF (MSSM) curve peaks slightly to the left (right) of the
others in Process 1 (2) for all mass spectra, although to different degrees.  Also, the
FSS and FSV curves are very similar particularly at SPS 1a and 9.

From these curves for processes 1 and 2, we can construct the distribution of processes
through a $W^-$ and the distribution of processes through a $W^+$, which are the
distributions which would actually be observed.  If an $\ell^-$ is observed, the chain must have
started with the partner of either a down-type quark, or an up-type antiquark.  We define
$r_{d^*}=1-r_{\bar{u}^*}$ to be the fraction of chains with an $\ell^-$ which begin with
the partner of a down-type quark.  Similarly, we define $r_{u^*}=1-r_{\bar{d}^*}$ to
be the fraction of chains with an $\ell^+$ which begin with the partner of an up-type quark.
The $q\ell^{\mp}$ distributions, $\ud P_{\mp}/\ud \mhat$, are given by:
\begin{eqnarray}
  \label{eq:P+-}
  \frac{\ud P_-}{\ud \mhat} &=& r_{d^*} \frac{\ud P_1}{\ud \mhat} + r_{\bar{u}^*}
  \frac{\ud P_2}{\ud \mhat} \nonumber \\
  \frac{\ud P_+}{\ud \mhat} &=& r_{u^*} \frac{\ud P_2}{\ud \mhat} + r_{\bar{d}^*}
  \frac{\ud P_1}{\ud \mhat}.
\end{eqnarray}
No distinction between flavours of quarks was required in the earlier studies
\cite{Barr:2004ze,Smillie:2005ar,Datta:2005zs,Athanasiou:2006ef} of the cascade
decay of a quark partner.  In these the quark partner decayed straight into a quark and
a neutral particle so no charge information of the original quark was transmitted to the
rest of the chain making the results flavour independent.

\begin{figure}[!htbp]
  \centering
  \includegraphics[width=0.47\textwidth]{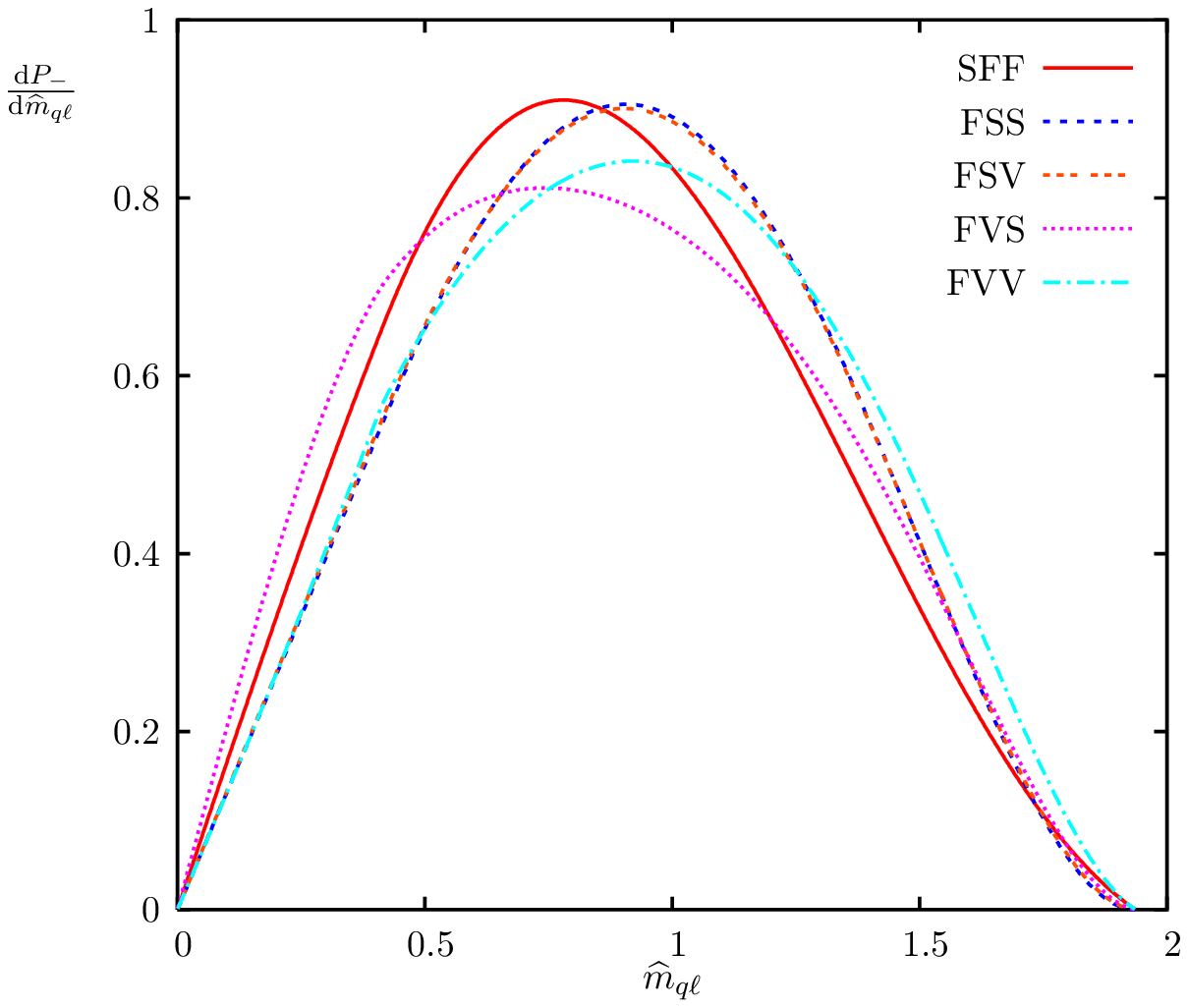}
  \includegraphics[width=0.47\textwidth]{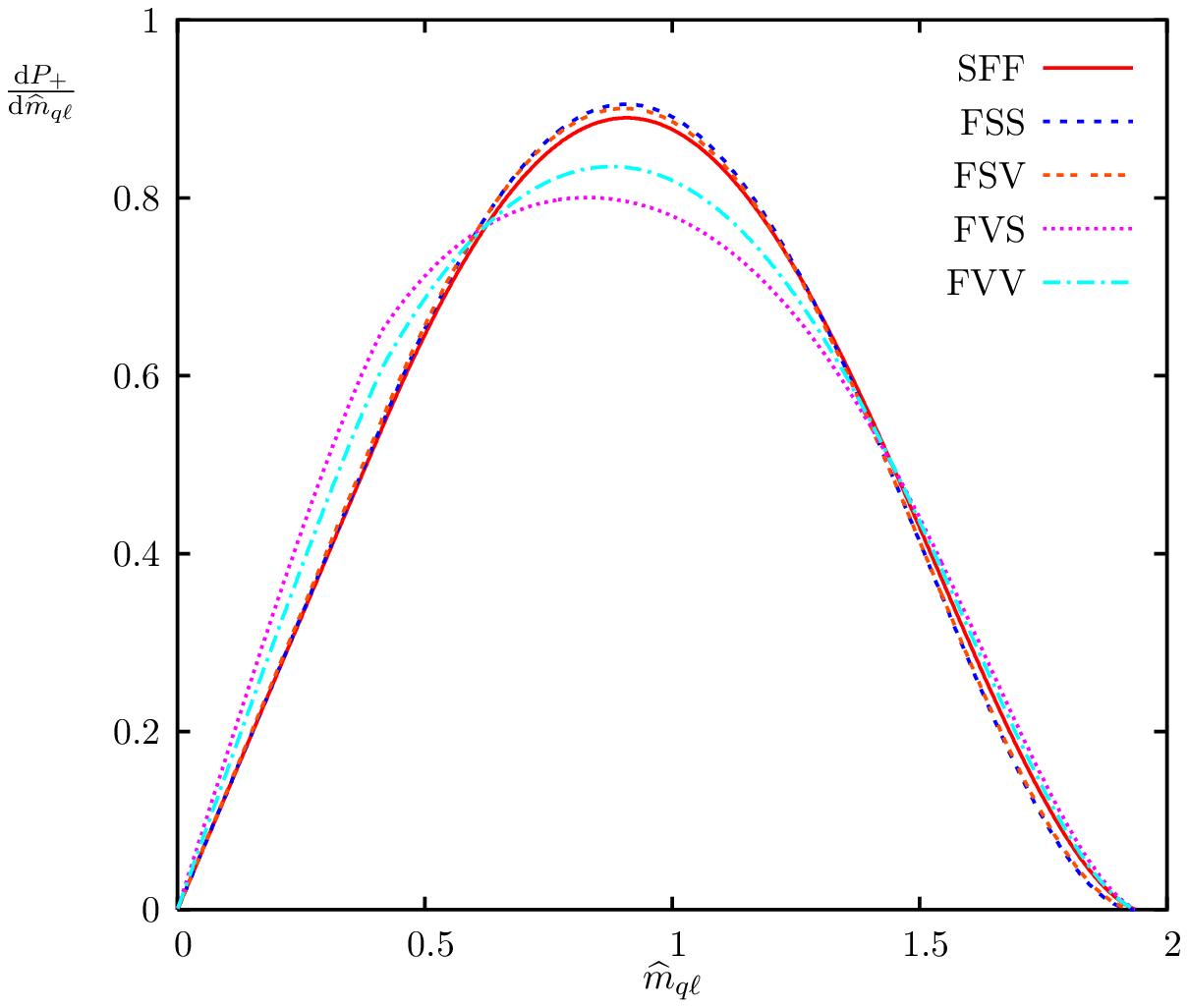}
  \caption{Observable invariant mass distributions for SPS 1a: $P_-$ (top) and $P_+$
    (bottom), see equation (\ref{eq:P+-}).} 
  \label{fig:PMandPP}
\end{figure}
\begin{figure}[!htbp]
  \centering
  \includegraphics[width=0.47\textwidth]{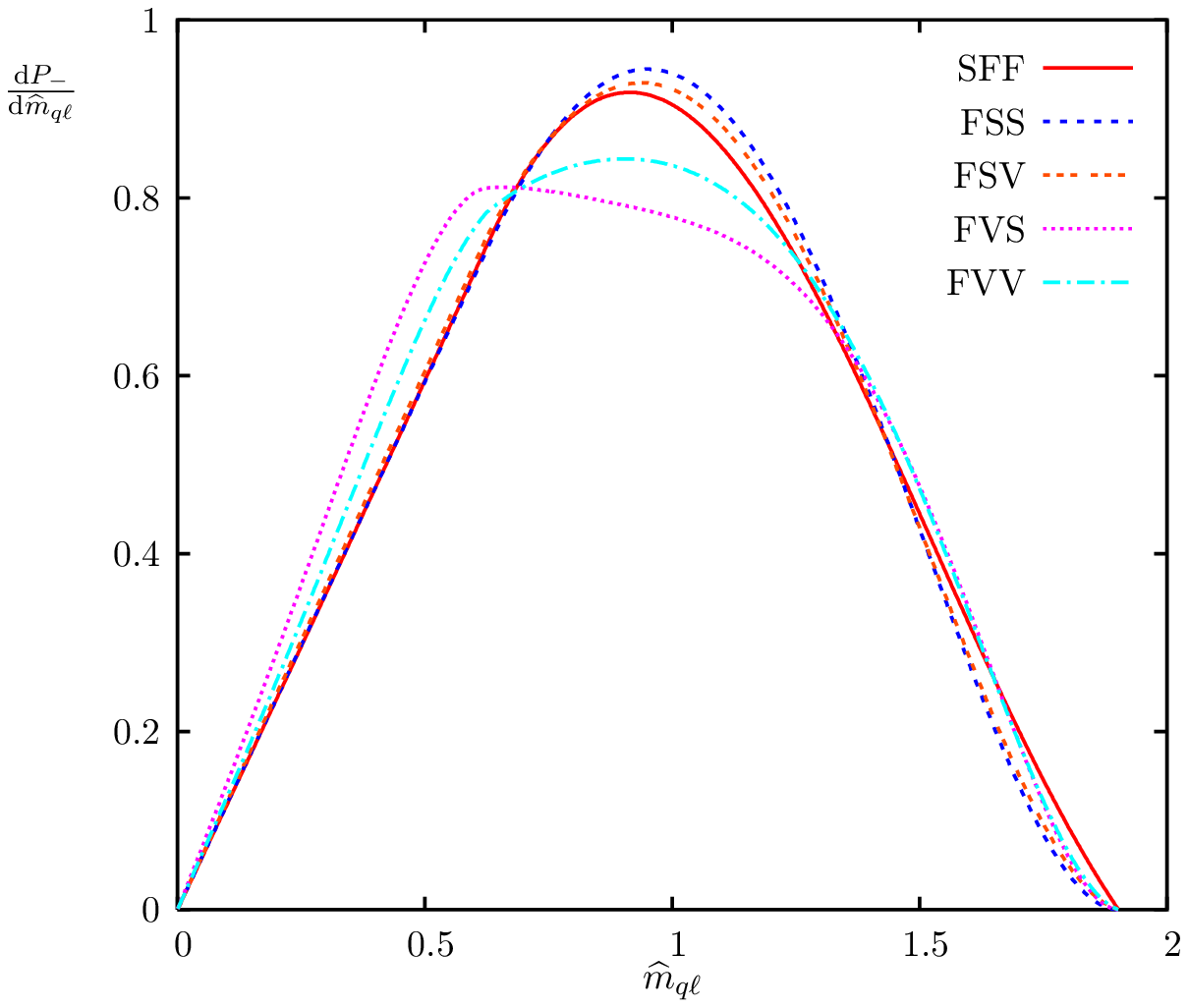}
  \includegraphics[width=0.47\textwidth]{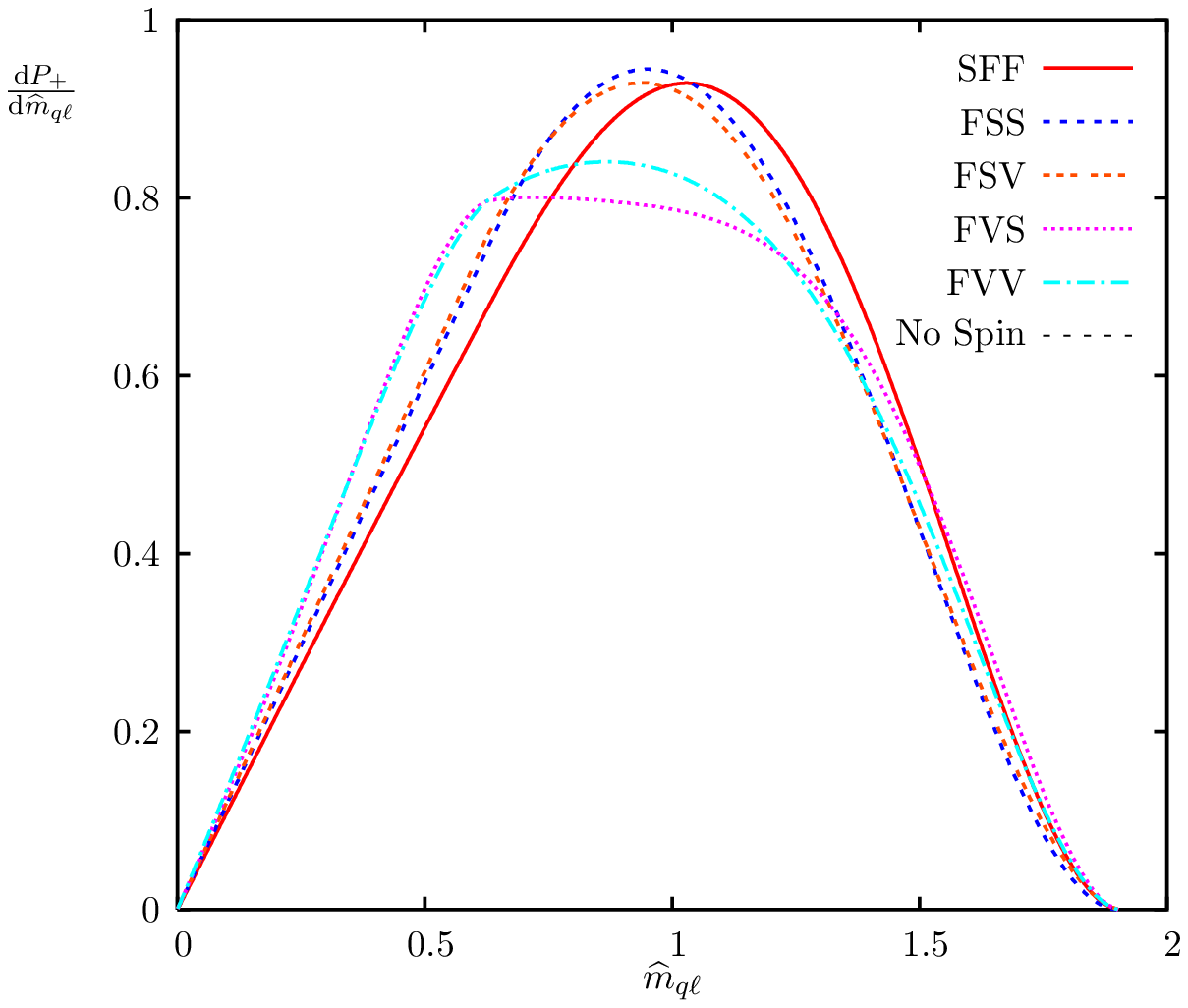}
  \caption{Observable invariant mass distributions for SPS 2: $P_-$ (top) and $P_+$
    (bottom)), see equation (\ref{eq:P+-}).} 
  \label{fig:SPS2PM}
\end{figure}
\begin{figure}[!htbp]
  \centering
  \includegraphics[width=0.47\textwidth]{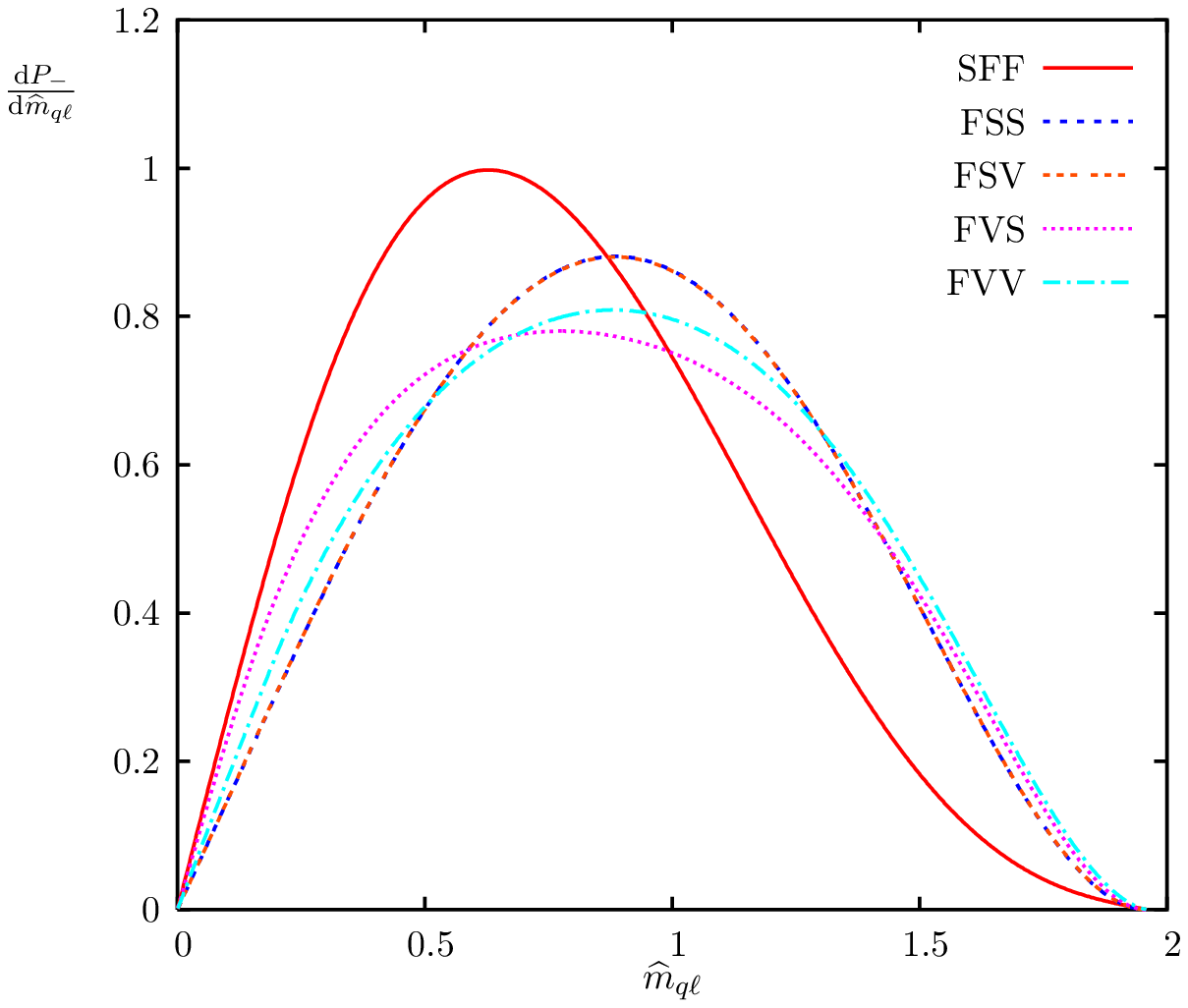}
  \includegraphics[width=0.47\textwidth]{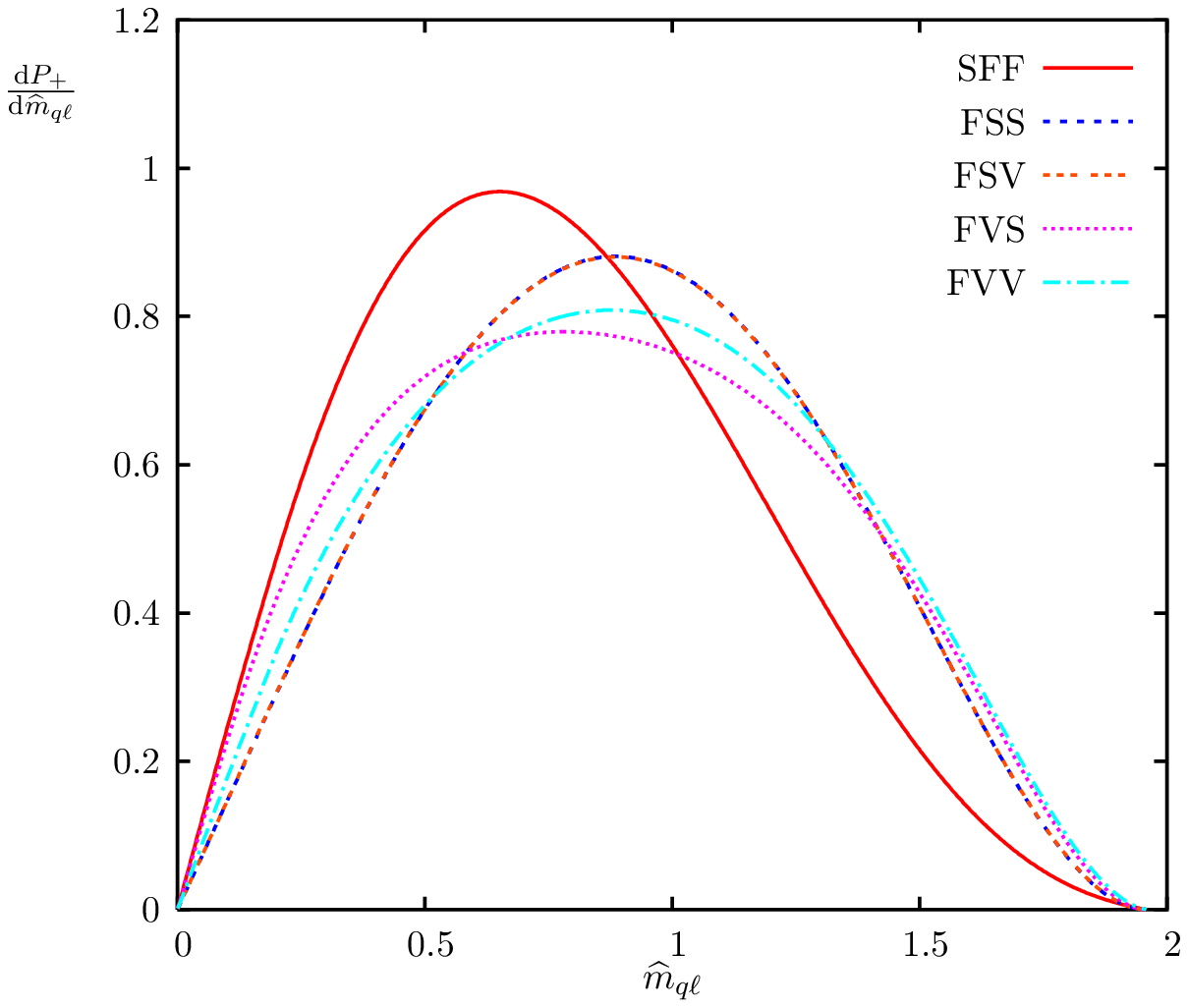}
  \caption{Observable invariant mass distributions for SPS 9: $P_-$ (top) and $P_+$
    (bottom)), see equation (\ref{eq:P+-}).} 
  \label{fig:SPS9PM}
\end{figure}

The MSSM scenarios in table \ref{tab:sps1a} imply the values of the fractions in table
\ref{tab:fractions} at the LHC, i.e. in $pp$ collisions at 14 TeV.\footnote{These results
  were obtained from the \textsc{Herwig}
  \cite{Corcella:2000bw,Corcella:2002jc,Moretti:2002eu} Monte Carlo at parton level,
  corresponding to the leading-order QCD production processes and MRST parton distribution
  functions \cite{Martin:1998np}.  They are not sensitive to details of the Monte Carlo,
  higher-order corrections or PDF uncertainties.}  Therefore these values represent
studying models with the MSSM flavour structure but different spin assignments.  We see
that at SPS 9, the effect of having more up quarks than down quarks in the proton is
dwarfed by the latter's larger branching ratio to a chargino.  This is caused by the large
value of the MSSM parameter $\mu$ enhancing the effect of large $\tan\beta$ in the
chargino mixing matrices.  The resulting plots are shown in figures \ref{fig:PMandPP}
--~\ref{fig:SPS9PM}.
\begin{table}[!htbp]
  \centering
  \begin{tabular}{|c|c|c|c|c|}
    \hline
    Spectrum & $r_{d^*}$ & $r_{\bar{u}^*}$ & $r_{u^*}$ & $r_{\bar{d}^*}$ \\ \hline
    SPS 1a & 0.860 & 0.140 & 0.469 & 0.531 \\ \hline
    SPS 2 & 0.900 & 0.100 & 0.911 & 0.089 \\ \hline
    SPS 9 & 0.998 & 0.002 & 0.072 & 0.928 \\ \hline
  \end{tabular}
  \caption{Numerical calculation of fractions using \textsc{Herwig}.}
  \label{tab:fractions}
\end{table}

We see that at SPS 9 the plots are nearly identical due to the extreme values of $r_{d^*}$
and $r_{\bar{d}^*}$ there.  There is greater variation in the individual curves at SPS 9
than for the other two mass spectra.  Our ability to distinguish the curves is discussed
in section \ref{sec:model-discrimination}.

\begin{figure}[!htbp]
  \centering
  \includegraphics[width=0.47\textwidth]{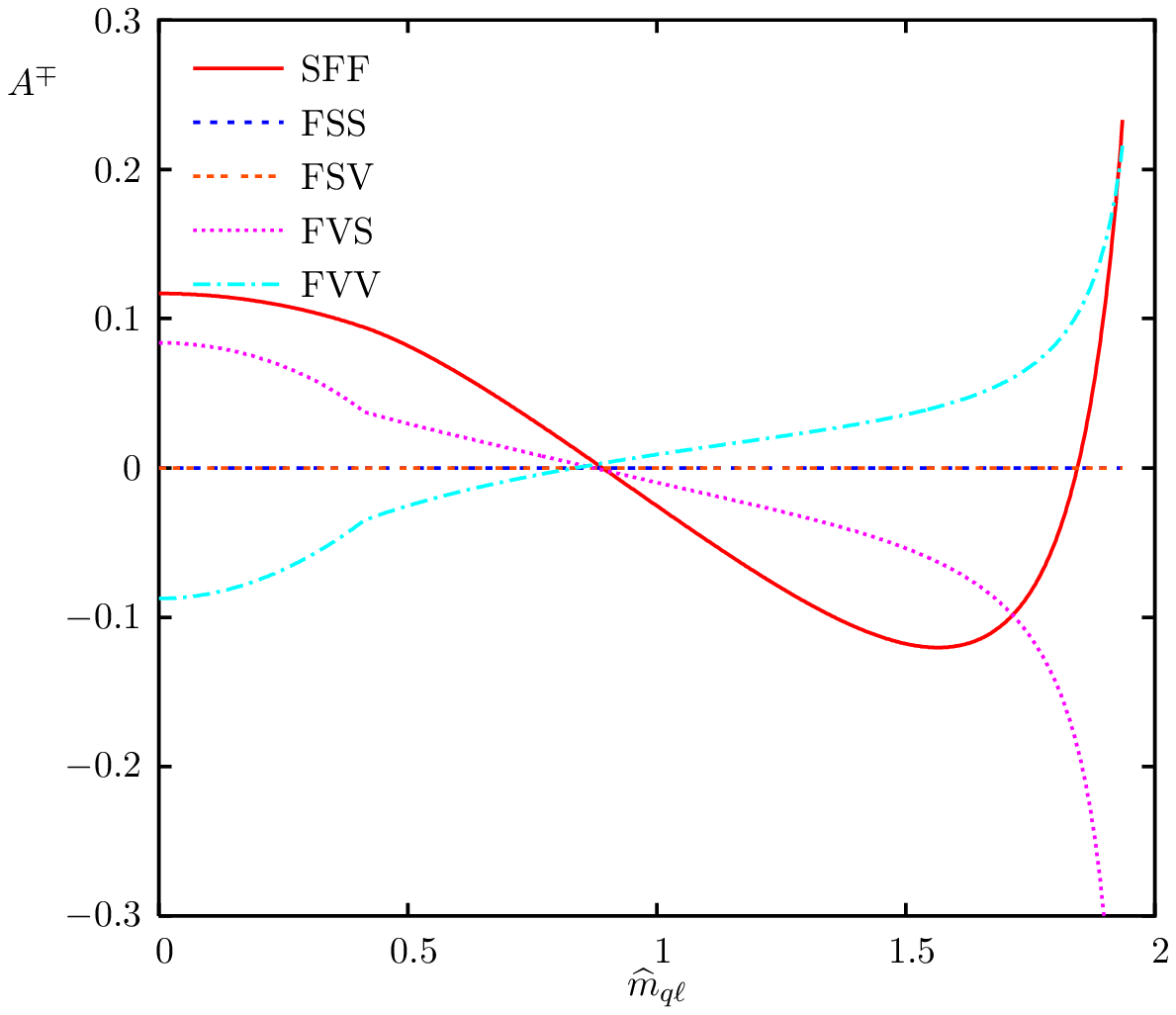}
  \includegraphics[width=0.47\textwidth]{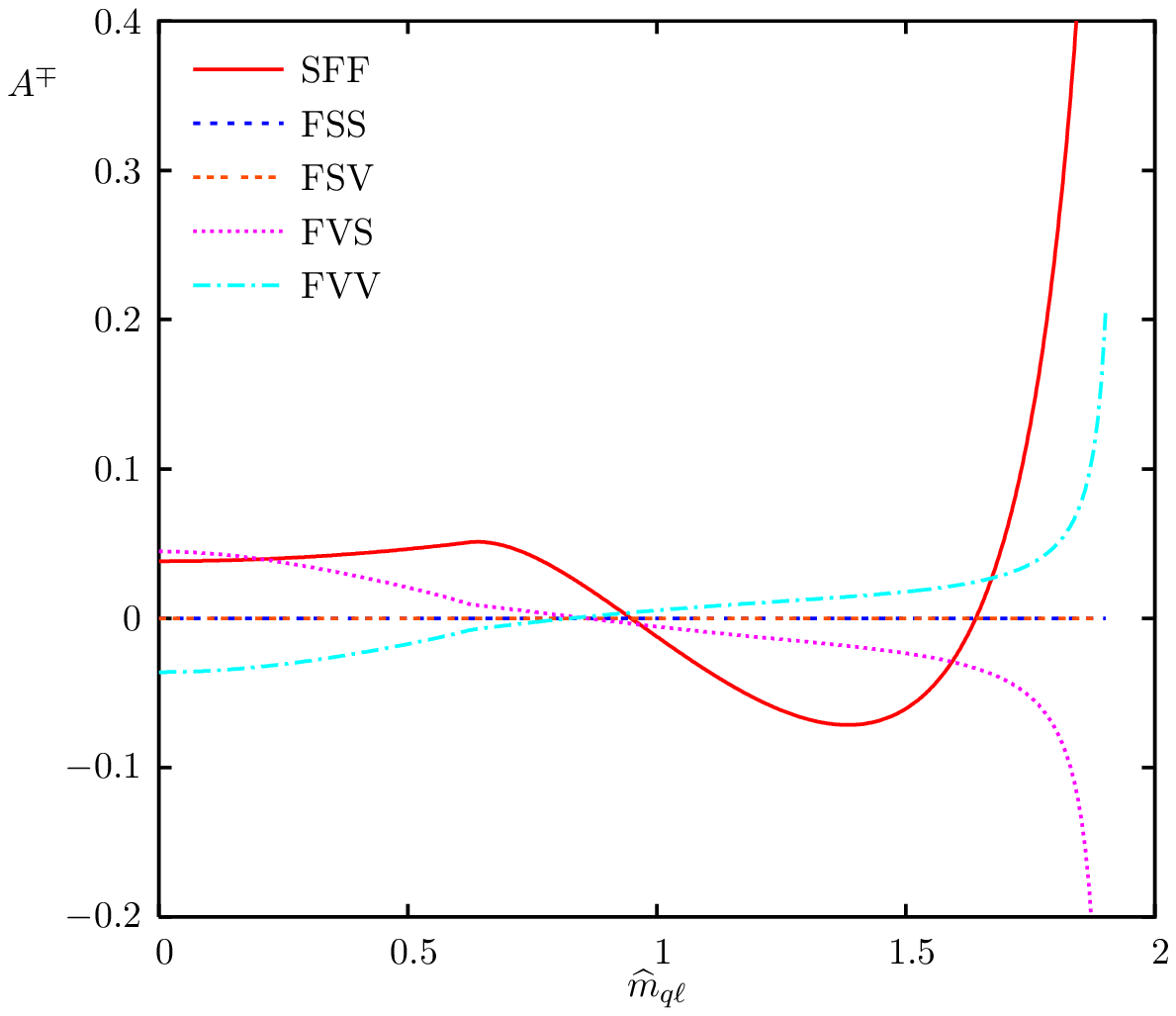}
  \includegraphics[width=0.47\textwidth]{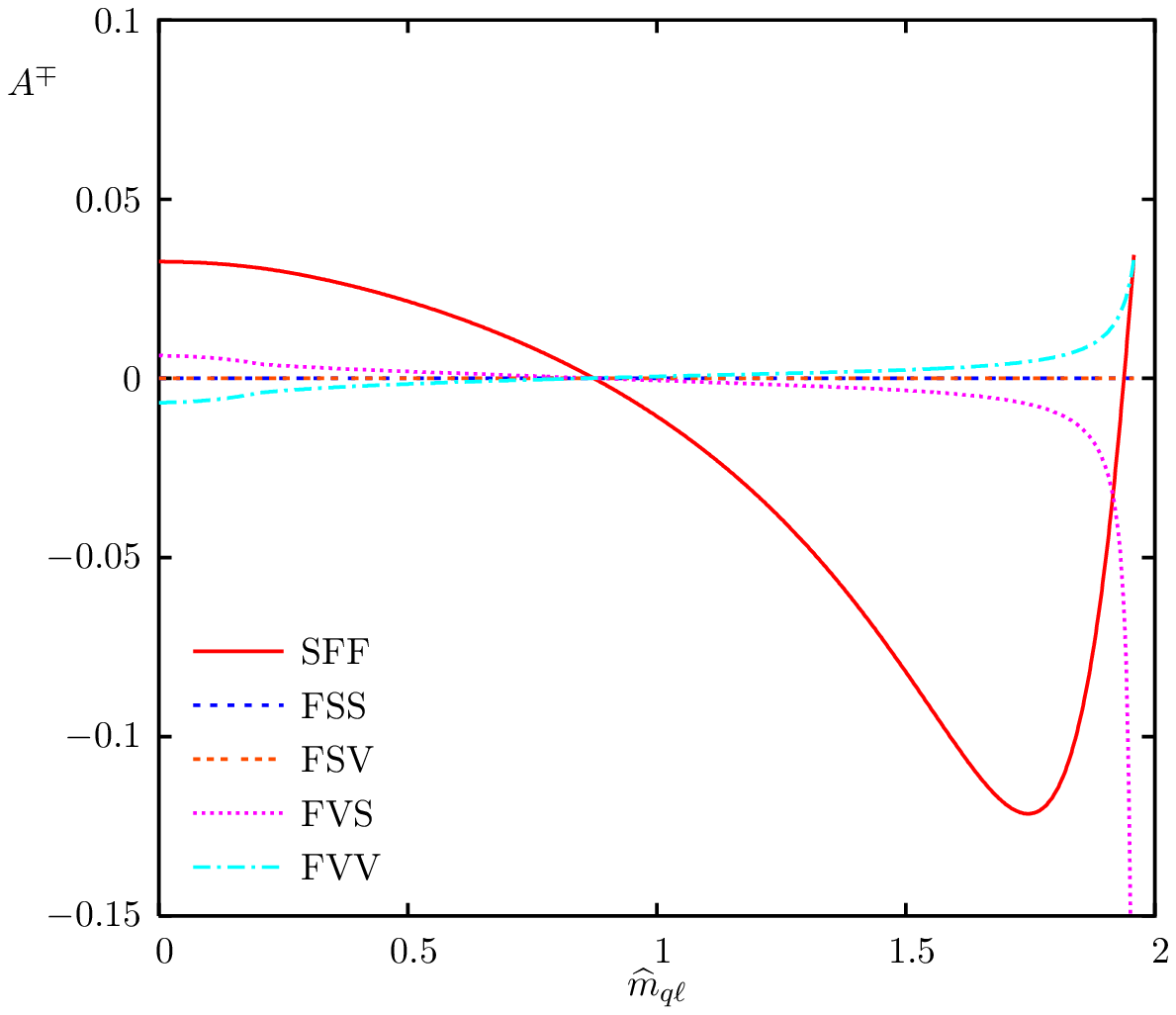}
  \caption{Asymmetry plots for SPS 1a (top), SPS 2 (middle) and SPS 9 (bottom).} 
  \label{fig:Asymm}
\end{figure}

We combine the information from both chains together by forming the asymmetry
of the normalised distributions given by
\begin{eqnarray}
  \label{eq:asymm}
  A^{\mp} = \frac{\phantom{d}\frac{\ud P_-}{\ud \mhat^2} - \frac{\ud P_+}{\ud
      \mhat^2}\phantom{d}}{\frac{\ud P_-}{\ud \mhat^2} + \frac{\ud P_+}{\ud \mhat^2}}.
\end{eqnarray}
The resulting plots are given in figure \ref{fig:Asymm}.

The asymmetry at SPS 1a (figure~\ref{fig:Asymm}, top) shows a difference in the behaviour
of the SFF, FVS and FVV curves.  With a 10\% level of asymmetry, we can be optimistic
about distinguishing the SFF (MSSM) curve. The differences at very high and low $\mhat$
cannot usually be used as this is where experimental statistics are often much worse.

In the asymmetry plot for SPS 2 (figure~\ref{fig:Asymm}, middle), it is unlikely that we
would be able to distinguish the FVS and FVV curves from the FSS and FSV line of zero
asymmetry.  The SFF line peaks at below 10\% making it also difficult to observe.

The same plot for SPS 9 (figure~\ref{fig:Asymm}, bottom) shows low levels of asymmetry
for all the curves, with the exception of the SFF curve.  Its peak of over 10\% asymmetry
suggests that for these masses it could be picked out.  It is unlikely that any of the
other curves could be distinguished from each other for any of these mass spectra.

\clearpage

\section{Model discrimination}
\label{sec:model-discrimination}

Here we apply the Kullback-Leibler distance \cite{Kullback:1951}. In our notation, it is
defined as
\begin{equation}
  \mathrm{KL}(T,S)=\int_m\log\left(\frac{p(m|T)}{p(m|S)}\right) p(m|T) \, \ud m ,
  \label{eq:KL}
\end{equation}
where $p(m|T)$ is the probability density function for $m$ given the distribution $T$,
and analogously for $p(m|S)$.  The expression that distribution $T$ is 
$R$ times more likely than distribution $S$, on the basis of the data points $\{m_i\}$, is
\begin{equation}
  \label{eq:Rdef}
  R=\frac{p(T|\{m_i\})}{p(S|\{m_i\})}.
\end{equation}
It was shown in \cite{Athanasiou:2006ef} that this can be rearranged to give a minimum
number of events, $N$, needed such that distribution $T$ is calculated to be $R$ times
more likely than distribution $S$, assuming that $T$ is the true distribution: 
\begin{equation}
  \label{eq:N}
  N\sim \frac{\log R + \log \frac{p(S)}{p(T)}}{\mathrm{KL}(T,S)}
\end{equation}
in the limit $N \gg 1$ where $p(S)$ and $p(T)$ are the prior probabilities of each
distribution.  We must make an assumption about the true distribution as we must generate
our data points for comparison from a particular distribution.  This will not be the case
when we have real data.  We include the factor $\log (${\small $p(S)/p(T)$}$)$ for
completeness, however, we will set it to zero in our analysis.  This is equivalent to
assuming all distributions to be equally likely before we look at the data.  Also, as
pointed out in \cite{Athanasiou:2006ef}, the result is invariant under diffeomorphisms
$m\to f(m)$, so the result would be unaffected if we had calculated with functions of
$\mhat^2$ for example, instead of $\mhat$.

The value $N$ is an absolute lower bound on the required number of events.  Once
background and detector effects are included these will rise considerably, however, these
effects vary from experiment to experiment and hence it is useful to have a universal
lower bound.

The results for the observable $P_{\mp}$ distributions at SPS 1a
(figure~\ref{fig:PMandPP}) are given in table~\ref{tab:KLresults}.  The corresponding
results for the curves at SPS 2 (figure~\ref{fig:SPS2PM}) and SPS 9
(figure~\ref{fig:SPS9PM}) are shown in tables \ref{tab:KL-SPS2} and \ref{tab:KL-SPS9}.
The value $R=1000$ has been taken here, so we are asking for one model to appear 1000
times more likely than another.  This corresponds to a 99.9\% confidence level, but is a
matter of choice.

\begin{table}[b]
\centering
\begin{tabular}[]{@{}r@{$\;$}|@{$\;$}r@{$\;$}r@{$\;$}r@{$\;$}r@{$\;$}r@{}}
(a) & SFF & FSS & FSV & FVS & FVV  \\
\hline
SFF  & $\infty$ & 697 & 756 & 1237 & 468 \\
FSS & 712 & $\infty$ & 177273 & 464 & 1481 \\
FSV & 764 & 171080 & $\infty$ & 498 & 1687 \\
FVS & 1221 & 434 & 466 & $\infty$ & 444 \\
FVV & 448 & 1298 & 1512 & 465 & $\infty$ \\
\end{tabular}

\vspace{0.3cm}
\begin{tabular}[]{@{}r@{$\;$}|@{$\;$}r@{$\;$}r@{$\;$}r@{$\;$}r@{$\;$}r@{}}
(b) & SFF & FSS & FSV & FVS & FVV  \\
\hline
SFF & $\infty$ & 6728 & 9459 & 975 & 2801 \\
FSS & 7728 & $\infty$ & 177273 & 732 & 1689 \\
FSV & 10408 & 171080 & $\infty$ & 819 & 2022 \\
FVS & 938 & 688 & 778 & $\infty$ & 5523 \\
FVV & 2734 & 1590 & 1932 & 5605 & $\infty$ \\
\end{tabular}
\caption{Number of events needed, with SPS 1a masses, to disfavour the column model with
  respect to the row model by a factor of 1/1000, assuming the data to come from the row
  model, for (a) $\ud P_-/\ud \mhat$ and (b) $\ud P_+/\ud \mhat$.}
\label{tab:KLresults}
\end{table}

\begin{table}
\centering
\begin{tabular}[]{@{}r@{$\;$}|@{$\;$}r@{$\;$}r@{$\;$}r@{$\;$}r@{$\;$}r@{}}
(a) & SFF & FSS & FSV & FVS & FVV  \\
\hline
SFF & $\infty$ & 1220 & 2223 & 704 & 2166 \\
FSS & 1608 & $\infty$ & 19314 & 570 & 1292 \\
FSV & 2668 & 17780 & $\infty$ & 738 & 2047 \\
FVS & 721 & 560 & 730 & $\infty$ & 3181 \\
FVV & 2267 & 1240 & 2016 & 3211 & $\infty$ \\
\end{tabular}

\vspace{0.3cm}
\begin{tabular}[]{@{}r@{$\;$}|@{$\;$}r@{$\;$}r@{$\;$}r@{$\;$}r@{$\;$}r@{}}
(b) & SFF & FSS & FSV & FVS & FVV  \\
\hline
SFF & $\infty$ & 1484 & 1468 & 586 & 649 \\
FSS & 1531 & $\infty$ & 19314 & 639 & 1106 \\
FSV & 1483 & 17780 & $\infty$ & 853 & 1655 \\
FVS & 572 & 619 & 840 & $\infty$ & 5551 \\
FVV & 630 & 1081 & 1638 & 5660 & $\infty$ \\
\end{tabular}
\caption{As in table \ref{tab:KLresults} for the SPS 2 mass spectrum, for (a) $\ud P_-/\ud
  \mhat$ and (b) $\ud P_+/\ud \mhat$.}
\label{tab:KL-SPS2}
\end{table}

\begin{table}
\centering
\begin{tabular}[]{@{}r@{$\;$}|@{$\;$}r@{$\;$}r@{$\;$}r@{$\;$}r@{$\;$}r@{}}
(a) & SFF & FSS & FSV & FVS & FVV  \\
\hline
SFF & $\infty$ & 90 & 90 & 118 & 87 \\
FSS & 83 & $\infty$ & 5939353 & 648 & 1686 \\
FSV & 83 & 5888890 & $\infty$ & 659 & 1734 \\
FVS & 97 & 608 & 618 & $\infty$ & 1780 \\
FVV & 73 & 1605 & 1654 & 1844 & $\infty$ \\
\end{tabular}

\vspace{0.3cm}
\begin{tabular}[]{@{}r@{$\;$}|@{$\;$}r@{$\;$}r@{$\;$}r@{$\;$}r@{$\;$}r@{}}
(b) & SFF & FSS & FSV & FVS & FVV  \\
\hline
SFF & $\infty$ & 123 & 124 & 162 & 121 \\
FSS & 117 & $\infty$ & 5939353 & 666 & 1686 \\
FSV & 118 & 5888890 & $\infty$ & 677 & 1735 \\
FVS & 139 & 626 & 637 & $\infty$ & 2176 \\
FVV & 105 & 1609 & 1659 & 2253 & $\infty$ \\
\end{tabular}
\caption{As in table \ref{tab:KLresults} for the SPS 9 mass spectrum, for (a) $\ud
  P_-/\ud\mhat$ and (b) $\ud P_+/\ud \mhat$.}
\label{tab:KL-SPS9}
\end{table}
 
The lower numbers in the SPS 9 tables reflect the original impression from the graphs that
the curves are easier to separate at this point than at SPS 1a or 2.  The exception is
between the FSS and FSV curves, which was to be expected from the similar functional form.
The values for SFF (which corresponds to the MSSM) are the lowest, but are still of the
order of 100.  These will be degraded in an experimental situation.

The numbers in tables \ref{tab:KLresults}~--~\ref{tab:KL-SPS9} treat the $W^+$ and $W^-$
chains individually, however, we can reasonably expect that if one is observed, both will
be.  The relative numbers of the two chains again depend on the masses in the chain.  With
experimental data these values would be known, but here we rely on the MSSM values
obtained in the same way as those in table \ref{tab:fractions}.  The results are shown in table \ref{tab:fpm}, where the fraction of $W$
chains which included a $W^{\pm}$ is denoted $f_{\pm}$.

When we consider both sets of data at once, equation (\ref{eq:KL}) is generalised to
\begin{eqnarray}
  \label{eq:BigKL}
  {\rm KL}(T,S) \; &=& \; \int_m \log \left( \frac{p(m^+|T^+)}{p(m^+|S^+)} \right)
  p(m^+|T^+) \nonumber \\ && \quad + \log\left( \frac{p(m^-|T^-)}{p(m^-|S^-)} \right) p(m^-|T^-) \; \ud m 
  \nonumber  \\
  &=& \quad \: \: {\rm KL}^+(T,S) \quad + \quad {\rm KL}^-(T,S)
\end{eqnarray}
where $p(m^{\pm}|U^{\pm})=f_{\pm} \; p(m|U^{\pm})$.  This gives the values of $N$ shown in
table \ref{tab:Comb1a2}.  
\begin{table}[!htbp]
  \centering
  \begin{tabular}{|c|c|c|}
    \hline
    Spectrum & $f_+$ & $f_-$ \\
    \hline
    SPS 1a & 0.57 & 0.43 \\
    SPS 2 & 0.68 & 0.32 \\
    SPS 9 & 0.67 & 0.33 \\
    \hline
  \end{tabular}
  \caption{Fractions, $f_{\pm}$, of total number of $W$ chains which include a $W^{\pm}$
    for each mass spectrum.}
  \label{tab:fpm}
\end{table}

\begin{table}[!t]
\centering
\begin{tabular}[]{@{}r@{$\;$}|@{$\;$}r@{$\;$}r@{$\;$}r@{$\;$}r@{$\;$}r@{}}
 (a) & SFF & FSS & FSV & FVS & FVV  \\
\hline
SFF  & $\infty$ & 1425 & 1589 & 1073 & 891 \\
FSS & 1476 & $\infty$ & 1.8$\cdot 10^5$ & 587 & 1593 \\
FSV & 1619 & 1.8$\cdot 10^5$ & $\infty$ & 642 & 1863 \\
FVS & 1041 & 549 & 604 & $\infty$ & 933 \\
FVV & 855 & 1450 & 1726 & 975 & $\infty$ \\
\end{tabular}

\vspace{0.3cm}
\begin{tabular}[]{@{}r@{$\;$}|@{$\;$}r@{$\;$}r@{$\;$}r@{$\;$}r@{$\;$}r@{}}
 (b) & SFF & FSS & FSV & FVS & FVV  \\
\hline
SFF  & $\infty$ & 1388 & 1647 & 619 & 837 \\
FSS & 1554 & $\infty$ & 1.9$\cdot 10^4$ & 615 & 1160 \\
FSV & 1729 & 1.8$\cdot 10^4$ & $\infty$ & 812 & 1763 \\
FVS & 613 & 599 & 801 & $\infty$ & 4482 \\
FVV & 819 & 1127 & 1742 & 4550 & $\infty$ \\
\end{tabular}

\vspace{0.3cm}
\begin{tabular}[]{@{}r@{$\;$}|@{$\;$}r@{$\;$}r@{$\;$}r@{$\;$}r@{$\;$}r@{}}
 (c) & SFF & FSS & FSV & FVS & FVV  \\
\hline
SFF  & $\infty$ & 110 & 110 & 144 & 107 \\
FSS & 103 & $\infty$ & 5.9$\cdot 10^6$ & 660 & 1686 \\
FSV & 103 & 5.9$\cdot 10^6$ & $\infty$ & 671 & 1734 \\
FVS & 122 & 620 & 631 & $\infty$ & 2027 \\
FVV & 92 & 1607 & 1657 & 2100 & $\infty$ \\
\end{tabular}
\caption{Total number of $W^+$ and $W^-$ events needed to disfavour the column model with
  respect to the row model by factor of 1/1000, assuming data to come from the row model at
  (a) SPS 1a, (b) SPS 2 and (c) SPS 9.}
\label{tab:Comb1a2}
\end{table}


In order to illustrate how these numbers show an improvement over treating the
distributions separately, we consider the specific values of the (SFF,FVS) entry at SPS 9.
For the $P_-$ distribution alone it was 118, while for $P_+$ alone it was 162.  For this
mass spectrum, a third of the chains have a $W^-$.  This means that looking at the whole
sample, for every $W^-$ event there are roughly two $W^+$ events.  If the $W^+$ events
contributed no discriminatory information (that is if $p(m^+|{\rm SFF})=p(m^+|{\rm FVS})$
for all $m^+$), then we would expect to need about $3$ times the number of $W^-$
events alone, 354.  However, as the $W^+$ events do contribute to distinguishing the two
models, we find that only 144 events in total are required.

Equation (\ref{eq:N}) shows that when the prior probabilities of the models are equal
(i.e. $p(S)=p(T)$ as we have used) the number of events $N_1$ calculated for a
discrimination level $R=R_1$ is related to the number of events $N_2$ calculated with
$R=R_2$ by a multiplicative factor.  For example, to obtain the results for $R=20$ (which
corresponds to a 95\% confidence level), the numbers
in table \ref{tab:KLresults} should be multiplied by $\log(20)/\log(1000)\simeq 0.43$.

It is instructive to consider how these events translate into required luminosity.  The
cross sections for these chains in the MSSM are given in table \ref{tab:sigmas}, where the
branching ratios of $\chi^{\pm}_2 \to W$ and $W \to e,\mu$ have been taken into account.
This corresponds to considering the first row in each table --- that in which the MSSM is
the true scenario.  The quoted required luminosity is calculated using the maximum number
which appears in the first row of each table.

\begin{table}[htbp]
  \centering
  \begin{tabular}{|c|c|c|}
    \hline Spectrum & Cross Section (fb) & Luminosity (fb$^{-1}$) \\ \hline
    SPS 1a & 12.3 & 129 \\
    SPS 2 & 1.41 & 1171 \\
    SPS 9 & 0.03 & 5473 \\ \hline
  \end{tabular}
  \caption{Cross sections for chains of the form shown in figure \ref{fig:decaychain} in
    the MSSM and corresponding integrated luminosity.}
  \label{tab:sigmas}
\end{table}
The highest cross section is for the SPS 1a mass spectrum, which is as expected as it is
relatively light.  This gives a required luminosity of 129 fb$^{-1}$.  The design
integrated luminosity for the LHC is 300 fb$^{-1}$, before upgrade.  This is encouraging,
however, the required value will inevitably increase when detector and background effects
are considered.  It looks unlikely that these studies could be conducted at this level of
discrimination for either SPS 2 or SPS 9.  The effect of the low numbers at SPS 9 has been
suppressed by the small branching ratio of $\widetilde{q}_L \to q_L^{\prime}
\widetilde{\chi}^{\pm}_2$ at this point.

\section{Conclusions}
\label{sec:conclusions}

The spin correlations in the decay of a quark partner via a leptonic $W$ boson decay, as
exhibited in the invariant mass distributions of the quark and charged lepton, have been
studied for three distinct SUSY-inspired mass spectra (SPS points 1a, 2 and 9).  We have
considered the 5 possible spin assignments in the chain and studied the extent to which
they can be distinguished.  The observable invariant mass distributions were constructed
where we found that the distributions had similar functional form.  The asymmetry
constructed from these plots could be useful for distinguishing the SFF curve (which
corresponds to the MSSM) from any of the other curves, but depends on the mass spectrum.

The results were quantified using the Kullback-Leibler distance to give a lower bound on
the number of events required to distinguish the spin assignments at a given level of
certainty.  This was applied to the distributions individually, and then to them combined.
These provide a guide to how useful particular channels would be in a study like this.
The lowest numbers were for the SPS 9 mass spectrum where the lower bound was of the order
of 100 events when attempting to distinguish the SFF curve from others, and higher (in
some cases considerably) when attempting to distinguish between the other distributions.
It therefore seems that this could be a useful method to distinguish the MSSM from other
spin assignments in the chain, but will be less useful to distinguish amongst these
alternatives.

These bounds were converted to a luminosity requirement for the case where the MSSM was
the true scenario.  The values for the SPS 1a mass spectrum were encouraging, while for
SPS 2 and SPS 9 it appears unlikely that this method could give a high level of
discrimination.

\section*{Acknowledgements}
\label{sec:acknowledgements}

I thank members of the Cambridge Supersymmetry Working Group for helpful discussions while
this work progressed.  I am particularly grateful to Bryan Webber for constructive
comments, numerical checks and useful discussions throughout.  I also thank Jeppe Andersen
for technical assistance and Ben Allanach for clearing up some points and commenting on a
draft.

\appendix

\section{Analytical formulae}
\label{sec:analytical-formulae}

This section contains the formulae for the distributions plotted in section
\ref{sec:spin-correlations} (with the exception of FVV, see below).  They are
expressed in terms of two constants $k_1$ and $k_2$ which are functions of the mass ratios
described in section \ref{sec:spin-correlations}:
\begin{equation}
  \label{eq:k12}
  k_1=1+Y-Z \qquad \mathrm{and} \qquad k_2=\sqrt{k_1^2-4Y},
\end{equation}
such that $k_1 > k_2 > 0$.  Then equations (\ref{eq:mqmu}) and (\ref{eq:mhat}) give
\begin{eqnarray}
  \label{eq:mhatk12}
  \widehat{m}^2_{q\ell} &=& k_1(1-\cos\theta\cos\psi) + k_2(\cos\theta-\cos\psi) \nonumber
  \\ && \quad -
  2\sqrt{Y}\sin\theta\sin\psi\cos\phi
\end{eqnarray}
with maximum $2(k_1+k_2)$.  We define the shorthands $k_{12}^{\pm}=k_1\pm k_2$ and
$\mhat=\mhat_{q\ell}$.

Each distribution has different behaviour in the regions $0 \le \mhat^2 \le 2k_{12}^-$ and
$2k_{12}^- \le \mhat^2 \le 2k_{12}^+$, as can be seen in figures \ref{fig:P1and2} --
\ref{fig:SPS9} and in the equations below. This is because high values of $\mhat^2$ can
only occur for very specific angle configurations which cuts down the phase space and
leads to logarithmic behaviour.  This can be seen in the ``No spin'' curve in figures
\ref{fig:P1and2}(top), \ref{fig:SPS2}(top) and \ref{fig:SPS9}(top) which represents the
phase space.  Without this effect, the ``No Spin'' distribution would continue linearly in
$\mhat^2 \ge 2k_{12}^-$.

\vspace{0.3cm}
\textbf{SFF}

In the MSSM, the structure of the $B$-$W$-$A$ vertex is $1+\alpha \gamma_5$ where $\alpha$
is defined by the parameters of the model.  As this varies at each mass point, it is left
explicitly in the equations below.  Table \ref{tab:alpha} lists the values of $\alpha$ at
the particular points studied in this paper.
\begin{table}[!htbp]
  \centering
  \begin{tabular}{|c|c|c|c|}
    \hline Mass Spectrum & SPS 1a & SPS 2 & SPS 9 \\
    \hline  $\alpha$ & 0.5083 & 0.3875 & 0.8155 \\ \hline
  \end{tabular}
  \caption{Numerical values of the parameter $\alpha$ for the different mass spectra
    studied in the text.}
  \label{tab:alpha}
\end{table}

\begin{eqnarray}
  \label{eq:SFF}
  \frac{\ud P_1}{\ud \widehat{m}} &=& \frac{3\mhat}{32k_2((1 + \alpha^2)(k_1^2 + 2Y - 3k_1Y) -
        6Y\sqrt{Z}(1 - \alpha^2))}  \nonumber \\ && \; \times\left\{
    \begin{array}{l@{}l}
     16\left(k_2((1 + \alpha)^2k_1 + 4(1 + \alpha)^2Y \phantom{\frac{k_1}{k_2}}\right.& \\ \quad +
     \mhat^2(1 + 4\alpha + \alpha^2 +
            (-1 + \alpha^2)\sqrt{Z})) & \\ \quad -
        (2\alpha k_1\mhat^2  + (2(1 + \alpha)^2(1 + k_1) & \\ \quad \left. + (1 + \alpha^2)\mhat^2)Y)
          \log{\left(\frac{k_1 + k_2}{k_1 - k_2}\right)} \right) & \\ \qquad \qquad \qquad
        \qquad \qquad \qquad 0\le
        \mhat^2 \le 2k_{12}^- \\
      & \\ 8k_1^2(1 + \alpha)^2 + (1 + (-6 + \alpha)\alpha)\mhat^4 \\ \quad +
        8k_1((1 + \alpha)^2(k_2 + 4Y) & \\ \quad + \mhat^2(-2 - \alpha(2 + \alpha(2 + \sqrt{Z})) +
            \sqrt{Z})) \\ \quad + 8\mhat^2(-2(1 + \alpha)^2Y +
          k_2(1 + 4\alpha + \alpha^2 & \\ \quad - (1 - \alpha^2)\sqrt{Z}))  +
        16Y((1 + \alpha)^2(5 + 2k_2) & \\ \quad - 4(1 - \alpha^2)\sqrt{Z}) -
        16(2\alpha k_1\mhat^2 \\ \quad + (2(1 + \alpha)^2(1 + k_1) & \\ \quad + (1 + \alpha^2)\mhat^2)Y)
         \log{\left(\frac{2(k_1 + k_2)}{\mhat^2}\right)}
      & \\ \qquad \qquad \qquad
        \qquad \qquad \qquad 2k_{12}^- \le
      \mhat^2 \le 2k_{12}^+ & \\ 
    \end{array} \right. \\ \nonumber \\
  \frac{\ud P_2}{\ud \mhat} &=&  
  \frac{3\mhat}{32k_2((1 + \alpha^2)(k_1^2 + 2Y - 3k_1Y)-6Y\sqrt{Z}(1 - \alpha^2))}
  \nonumber \\ && \; 
  \times \left\{
    \begin{array}{l@{}l}
      -16\left(k_2(\mhat^2 + 4Y - k_1(1 - \alpha)^2 \phantom{\frac{k_1}{k_2}}\right. & \\ \quad +
          \alpha(\mhat^2(4 + \alpha) + 4\alpha Y)  -
          \mhat^2\sqrt{Z}(1 - \alpha^2)) \\ \quad - (Y(2 + \mhat^2 - 4\sqrt{Z})  +
          2\alpha(k_1\mhat^2 + 2Y) \\ \quad \left. + Y(2 + \mhat^2 + 4\sqrt{Z})\alpha^2)
         \log{\left(\frac{k_1 + k_2}{k_1 - k_2}\right)}\right) & \\ \qquad \qquad \qquad
        \qquad \qquad \qquad 0\le \mhat^2 \le 2k_{12}^- \\  & \\ 
       8k_1^2(1 - \alpha)^2 + \mhat^4(1 + \alpha(6 + \alpha)) & \\ \quad -
       16Y(3 + 2k_2 - 4\sqrt{Z} & \\ \quad +
         \alpha(10  + \alpha(3 + 2k_2 + 4\sqrt{Z})))& \\ \quad +
       8k_1(k_2(1 - \alpha)^2  + 2\mhat^2\alpha  -
         \mhat^2\sqrt{Z}(1 - \alpha^2) \\ \quad- 4Y(1 + \alpha^2))  -
       8\mhat^2(-2Y(1 + \alpha^2) \\ \quad + k_2(1 + 4\alpha +
           \alpha^2  + \sqrt{Z}(-1 + \alpha^2))) \\ \quad+
       16(Y(2 + \mhat^2 - 4\sqrt{Z}) + 2\alpha(k_1\mhat^2 + 2Y) \\ \quad +
         \alpha^2Y(2 + \mhat^2 + 4\sqrt{Z}))\log{\left(\frac{2(k_1 +
               k_2)}{\mhat^2}\right)} & \\ \qquad \qquad \qquad
        \qquad \qquad \qquad 2k_{12}^- \le \mhat^2 \le 2k_{12}^+ \\ 
    \end{array} \right.  
\end{eqnarray}

\textbf{FSS}

\begin{eqnarray}
  \label{eq:FSS}
  \frac{\ud P_{1,2}}{\ud \mhat} &=& \frac{3\mhat}{2k_2^3} \left\{
    \begin{array}{l l}
      k_1k_2-2Y\log\left(\frac{k_1+k_2}{k_1-k_2}\right) & \\ \qquad \qquad \qquad
      \qquad \qquad 0 \le \mhat^2 \le 2k_{12}^- \\
      & \\
      \frac{1}{16}(6k_1-2k_2-\mhat^2)(2(k_1+k_2)-\mhat^2) & \\ -2Y\log\left(\frac{2(k_1+k_2)}{\mhat^2}
      \right) & \\ \qquad \qquad \qquad \qquad \qquad 2k_{12}^- \le \mhat^2 \le 2k_{12}^+ \\
    \end{array} \right. 
\end{eqnarray}

\newpage
\textbf{FSV}

\begin{eqnarray}
  \label{eq:FSV}
  \frac{\ud P_{1,2}}{\ud \mhat} &=& \frac{3\mhat}{2k_2(k_2^2+12YZ)} \nonumber \\ && \times \left\{
    \begin{array}{l c}
      k_1k_2+2Y(2Z-1)\log \left(\frac{k_1+k_2}{k_1-k_2}\right) & \\ \qquad \qquad 
      \qquad \qquad \qquad 0\le \mhat^2\le
      2k_{12}^- \\
      & \\
      \frac{1}{16}(6k_1-2k_2-\mhat^2)(2(k_1+k_2)-\mhat^2) & \\
      \quad +2Y(2Z-1)\log\left(\frac{2(k_1+k_2)}{\mhat^2} \right) & \\ \qquad 
      \qquad \qquad \qquad \qquad 2k_{12}^-\le\mhat^2\le2k_{12}^+
    \end{array} \right.
\end{eqnarray}

\vspace{0.3cm}
\textbf{FVS}

For the FVS chain the parameter $a$ represents that in the
$C$-$q$-$B$ vertex, discussed at the end of section \ref{sec:spin-assignments}.


\begin{eqnarray}
  \frac{\ud P_1}{\ud \mhat} &=& \frac{9\mhat(k_1 + k_2)^2}{8k_2(1 + a^2)(1 + 2X)(k_1(k_1 +
    k_2) - 2Y)(k_1^2 + 8Y)} \nonumber \\ && \times \left\{
    \begin{array}{l c}
      2k_2(1 + a^2)(k_1 + 6\mhat^2) & \\ \quad -
      4Xk_2\mhat^2(3 + a(3a-2)) & \\ \quad +
      (\mhat^2(1 + a^2)(4k_1 + \mhat^2)(-1 + X) & \\ \quad -
      4Y(1 + a^2 - 2X(1 - a)^2))\log\left(\frac{k_1 + k_2}{k_1 - k_2}\right) & \\ \qquad
      \qquad \qquad \qquad \qquad \qquad  0 \le \mhat^2 \le 2k_{12}^- \\ & \\
      \frac{1}{8}\left(\left(2(2(k_1 + k_2) - \mhat^2)((1 + a^2)k_2(k_1 +
          k_2)\right. \phantom{\frac{k_1}{k_2}} \right.& \\ \quad \times
      (4k_2 + \mhat^2(15 - 16X)) & \\ \quad
      - 2(-(1 + a^2)(22k_1 + 26k_2 +
      15\mhat^2) & \\  \quad + 16((1 - a)^2(k_1 + k_2) & \\ \quad \left. + (1 + a^2)\mhat^2)X)
        Y)\right)/(k_1 + k_2)^2 & \\ \quad + 8((1 + a^2)\mhat^2(4k_1 + \mhat^2)(-1 + X) & \\
      \quad \left. -
        4Y(1 + a^2 - 2X(1 - a)^2))\log\left(\frac{2(k_1 + k_2)}{\mhat^2}\right)\right) & \\
      \qquad \qquad \qquad \qquad \qquad \qquad 2k_{12}^- \le \mhat^2 \le 2k_{12}^+ \\
    \end{array} \right.
\end{eqnarray}
  
\begin{eqnarray}
  \label{eq:FVS2}
  \frac{\ud P_2}{\ud \mhat} &=& \frac{\ud P_1}{\ud \mhat} \quad \mathrm{with} \quad a \to -a
\end{eqnarray}

\vspace{0.3cm}
\textbf{FVV}

The FVV distributions are too long to present here in a manageable way due to the
complicated $B^{\pm}$--$W^{\pm}$--$A$ vertex.  They are available on request from the
author.  They also have the symmetry (\ref{eq:FVS2}).

\section{Higher dimensional couplings}
\label{sec:loop-induc-coupl}

As mentioned in section \ref{sec:spin-assignments}, this analysis did not consider higher
dimensional couplings with a different Lorentz structure to that considered previously in
the paper.  For example, vertices with the form (a) $p.qg^{\mu\nu}-p^{\nu}q^{\mu}$ or (b)
$\epsilon^{\mu\nu\rho\sigma}p_{\rho}q_{\sigma}$ are studied in the context of anomalous
Higgs couplings \cite{Hankele:2006ma}.

\begin{figure}[!htbp]
  \centering
  \includegraphics[width=0.47\textwidth]{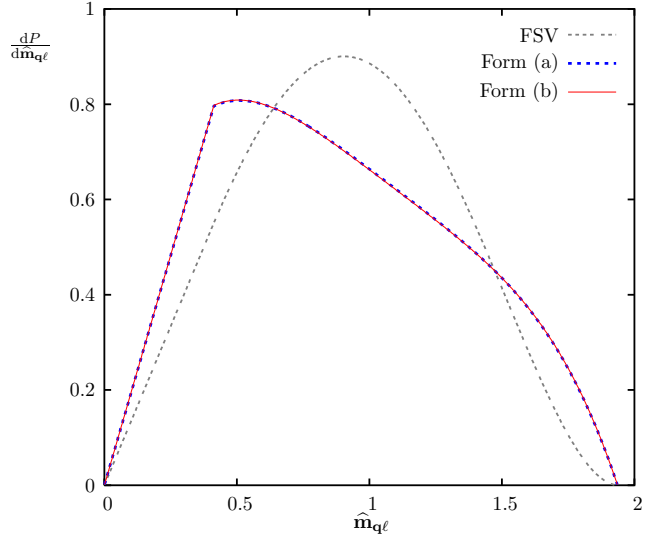}
  \setlength{\unitlength}{1cm}
  \caption{Distributions for the different scalar-vector-vector vertices described in the text.} 
  \label{fig:Even}
\end{figure}

Figure \ref{fig:Even} shows the distributions for the FSV chain with these different
vertices alongside the distribution shown before (marked FSV), for the SPS 1a
mass spectrum.

The new vertices give very similar distributions; the analytical expressions are given
below.  Due to the scalar in the chain the distributions for processes 1 and 2 are the same
and therefore so are the distributions for the chains with positive and negative leptons.
\begin{eqnarray}
  \label{eq:evenodd}
    \frac{\ud P_{\mathrm{(a)}}}{\ud \mhat} &=& \frac{3\mhat}{4k_2(k_2^2+6YZ)} \nonumber \\
    && \times \left\{
    \begin{array}{l c}
       -k_1k_2 +((k_1-2Y)^2+2Y)\log\left( \frac{k_1+k_2}{k_1-k_2}\right) & \\ \qquad
       \qquad \qquad \qquad \qquad \qquad 0\le \mhat^2\le 2k_{12}^- \\
      & \\
     \frac{1}{16} \left(-(6k_1-2k_2-\mhat)(2(k_1+k_2)-\mhat) \phantom{\frac{2(k_1)}{2}}
     \right. & \\ \left. \quad +16((k_1-2Y)^2+2Y)\log\left(\frac{2(k_1+k_2)}{\mhat^2}\right)
     \right)  & \\ \qquad
       \qquad \qquad \qquad \qquad \qquad 2k_{12}^-\le\mhat^2\le2k_{12}^+
    \end{array} \right.  \\ \phantom{a} \nonumber \\ 
   \frac{\ud P_{\mathrm{(b)}}}{\ud \mhat} &=& \frac{3\mhat}{4k_2^3} \left\{
    \begin{array}{l}
      -k_1k_2+(k_1^2-2Y)\log\left( \frac{k_1+k_2}{k_1-k_2}\right) \\ \qquad
       \qquad \qquad  \qquad 0\le \mhat^2\le 2k_{12}^- \\
      \\
      \frac{1}{16} \left(-(6k_1-2k_2-\mhat^2) \phantom{\frac{2(k_1)}{2}} \right. \\
      \quad \times (2(k_1+k_2)-\mhat^2) 
         \\
      \left. \quad +16(k_1^2-2Y)\log\left(\frac{2(k_1+k_2)}{\mhat^2}\right) \right) \\
      \qquad \qquad \qquad  \qquad 2k_{12}^-\le\mhat^2\le2k_{12}^+ 
    \end{array} \right. 
\end{eqnarray}

While the distributions are very similar to each other, they are quite different to that
of the same chain with the lowest order vertex and to the distributions for the other
chains shown in figure \ref{fig:PMandPP}.  Therefore vertices of this kind would be
unlikely to be mistaken for those already discussed.

\bibliography{Refs.bib}
\bibliographystyle{utphys}

\end{document}